\documentclass{jfm}
\usepackage{graphicx}
\usepackage{amsmath}
\usepackage{natbib}
\usepackage{xcolor}
\usepackage{hyperref}
\usepackage{wrapfig}
\usepackage{float}
\usepackage{xfrac}
\usepackage[utf8]{inputenc} % Enables UTF-8 support
\usepackage[T1]{fontenc}    % Proper font encoding for symbols
\usepackage{fixltx2e}
\usepackage{booktabs}
\usepackage{float}
\usepackage{textcomp}       % Extra symbols
\newcommand{\comment}[1]{}
\definecolor{dgr}{RGB}{0, 100, 0}

\newcommand{\upd}[1]{\textcolor{blue}{#1}}

\usepackage{epstopdf, epsfig}
\renewcommand{\b}[1]{\boldsymbol{#1}} 
\usepackage[percent]{overpic}

\shorttitle{Onset of unsteadiness in hypersonic SBLI on a cone-step}
\shortauthor{C. Jenquin, E. L. Cui, A. Dwivedi,
G. S. Sidharth and J. S. Jewell}

\title{ \Large Onset of separation unsteadiness in hypersonic shock boundary layer interaction on a cone-step}

\author{Chase Jenquin\aff{1}\corresp{\email{cjenquin@purdue.edu}}, Eric L. Cui\aff{1}, Anubhav Dwivedi\aff{2}, \newline G.S. Sidharth\aff{3} \and Joseph S. Jewell\aff{1}}

\affiliation{\aff{1}School of Aeronautics and Astronautics, Purdue University, West Lafayette, IN 47907, USA
\aff{2} Aerospace Engineering and Mechanics, University of Minnesota, 55455 , USA
\aff{3}Department of Aerospace Engineering, Iowa State University, Ames, IA 50011, USA}

\begin{document}

\maketitle

\begin{abstract}
Shock-boundary layer interactions (SBLI) on hypersonic cone step flows exhibit a range of intrinsic unsteady behaviors, from shear-layer oscillations to large-scale pulsations. This work investigates the unsteadiness in a cone-step geometry at Mach 6 under quiet flow conditions at different freestream Reynolds numbers using 
time-resolved Schlieren imaging and spectral proper orthogonal decomposition (SPOD). Experimental results are compared with high-fidelity axisymmetric and three-dimensional simulations. Results demonstrate regime transition in the parameter space, across the unsteadiness boundary, all the way from shear-layer breakdown to shock system oscillations and ultimately to large-amplitude pulsations. The dominant mode in the experiments and the simulations corresponds to a Strouhal number St $\approx 0.17$ for small oscillations reducing to St $ \approx 0.13$ for large pulsations. A detailed description of the unsteady shock dynamics, the instability of the shear layer during onset of unsteadiness and an analysis of the nonlinear limit cycle is presented. 
\end{abstract}


\section{Introduction}

\subsection{Flow Unsteadiness in Shock Boundary Layer Interactions}

Shock boundary layer interactions (SBLI) are ubiquitous in high-speed flow configurations \cite{babinsky2011}. A sudden adverse pressure gradient within the SBLI region can induce boundary layer separation, leading to unsteady flow phenomena across a wide range of temporal scales. This unsteadiness significantly impacts both thermal and aerodynamic loads, and understanding its origin is critical for high-speed flight vehicle design. Two primary sources contribute to the unsteady behavior of SBLI. First, the presence of  external disturbances, such as those associated with the  fluctuations within the incoming boundary layer or of an impinging shock can lead to unsteady SBLI response. Second, unsteadiness can also be triggered by intrinsic instabilities that arise from the inherent dynamics of the separation bubble itself, even in the absence of external perturbations. These instabilities can occur when the base-state reattaching shear layer is steady or even when it is unsteady due to Kelvin-Helmholtz \cite{} breakdown. This paper collectively refers to the latter as ``oscillation-type'' unsteadiness. Such unsteadiness observed in SBLIs has similarities to the oscillations observed in compressible flows over cavities where shear layers impinge on the corners.

%Flow unsteadiness in shock boundary layer interactions has been an important topic of research. Several factors contribute to unsteadiness: intrinsic instabilities and those induced from interaction with external disturbances, such as those in the boundary layer or an impinging shock. Intrinsic unsteadiness relates to the dynamics of the separation bubble in the absence of external disturbances. These can occur when the base-state reattaching shear layer is steady or when it is unsteady due to Kelvin-Helmholtz breakdown. This paper collectively refers to these as ``oscillation-type'' unsteadiness. The unsteadiness is also associated with oscillations observed in compressible flow over a cavity in which shear layers impinge on the corners.

%For geometries with spikes and large turn angle deviation for a significant mass of the incoming flow, a bow shock and a large separation zone exists. 
\upd{For geometries with spikes and large turn angles, a bow shock and a large separation zone exists.} In such flows unsteadiness may include shock oscillations where the separation shock foot moves along the length of the geometry. This phenomenon is generally known as ``pulsating'' unsteadiness. Pulsation unsteadiness has primarily been studied in the context of spiked axial-cylinders.  
\upd{It may be noted that while turbulent shock–boundary-layer interactions are also characterized by low-frequency breathing of the separation bubble \cite{clemens2014,touber2009large}, they have a relatively modest shock excursion and small separation lengths relative to the separating boundary layer thickness. In contrast, laminar separation lengths are significantly larger, and pulsations  exhibit unsteady separation zones that can span the entire forebody lengths.} 
%These oscillatory states, often classified as “pulsation” or “oscillation,” involve large-amplitude changes in separation-bubble size and topology and are controlled by a small set of geometric parameters, in contrast to the more amplifier-like response of canonical turbulent SBLI to incoming boundary-layer disturbances.

%Recently, both, oscillatory and pulsatory behavior has been observed in cone-step geometries. 
%which exhibit both axisymmetric shear layers and spike-like pulsatory behavior under certain conditions.
In this paper, we investigate a cone-step geometry where both oscillatory and pulsatory intrinsic modes exist.
%, at different freestream conditions. 
The study focuses on tracking the onset and evolution of the unsteadiness as the Reynolds number increases, transitioning from the oscillatory to the pulsatory state. The unsteadiness is shown to be a fluid-dynamic oscillation involving the global dynamics of the separation-reattachment shock system.

\section{Background}

% \sidgs{Citations}
% We provide a background of past studies relevant to oscillatory and pulsatory shock-boundary layer interactions, and the hypotheses that have developed over time to explain the unsteadiness. 

We present a comprehensive overview of prior investigations pertaining to oscillatory and pulsatory shock-boundary layer interactions.  We also provide a background on hypotheses for mechanisms driving unsteadiness in these interactions.

% We critically examine the evolution of theoretical frameworks and hypotheses that have been proposed to elucidate the mechanisms underlying the observed unsteadiness in such interactions.

\subsection{Oscillatory unsteadiness in reattaching shear layers}
\label{sec.osclitrev}

Oscillations in reattaching shear layers have been the subject of investigation since the work of \cite{rossiter1964report}, in which oscillations in open cavity flows were examined. \upd{It was established that deep rectangular cavities (with cavity length to depth ratio less than 4) are capable of supporting periodic unsteadiness due to acoustic resonance.} An empirical formula was introduced to predict the frequency of oscillations due to cavity resonance. This formulation is relevant to the present study, as a variant of it has recently been employed to characterize oscillatory unsteady disturbances in cone-step flows~\citep{Kumar_Sasidharan_Kumara_Duvvuri_2024}.

A comprehensive review on unsteadiness in reattaching shear layers was conducted by \cite{rockwell1978review}, where various cavity and related flow configurations exhibiting self-sustained oscillations were categorized. The interactions were classified into: (i) fluid-dynamic, associated with global instabilities induced by feedback mechanisms; (ii) fluid-resonant, driven by acoustic or surface wave phenomena; and (iii) fluid-elastic, in which wall structural responses influence the oscillatory behavior. It was noted that fluid-dynamic oscillations tend to dominate when the cavity length-to-acoustic-wavelength ratio is small. The upstream propagation of disturbances was identified as a key mechanism. Separately, when disturbances couple with standing wave resonance—such as that due to acoustic waves—the resulting behavior was designated as `fluid-resonant'.

To model the physical mechanism governing the oscillation cycle in shallow cavity fluid-resonant interactions, a pseudo-piston framework was proposed by \cite{heller1975piston}. Within this model, sinuous shear layer instabilities were shown to induce mass flux variations at the cavity trailing edge. The resulting unsteady shear layer motion was interpreted as generating a piston-like action, which in turn produces upstream-traveling waves that reinforce shear layer disturbances. The analytical model included internal wave structures comprising one-dimensional acoustic-like waves traveling in both upstream and downstream directions.

% \cite{heller1975piston} proposed a pseudo-piston type model to explain the physical mechanism of an oscillation cycle in fluid-resonant interactions in shallow cavities. Sinuous instabilities amplified in the shear layer result in mass addition and removal at the cavity trailing edge. The unsteady motion of the shear layer, therefore, sets up a piston-like effect that generates upstream traveling waves that reinforce the upstream traveling disturbances in the shear layer. In their analytic model, the internal wave structures in their physical model comprise upstream and downstream traveling one-dimensional acoustic-like waves.  

Later, observations of the three-dimensional flow structures in {shallow cavities} were analyzed using global stability analysis~\citep{Bres2008threedim}. It was identified that a three-dimensional recirculation zone characterized by a spanwise wavelength approximately equal to the cavity depth and oscillations occurring at frequencies roughly an order of magnitude lower than those of two-dimensional Rossiter-type (fluid-resonant or acoustic) instabilities~\citep{rockwell1978review}. The findings indicated that the dominant instability is {\it hydrodynamic rather than acoustic}, arising from a centrifugal instability mechanism associated with the mean recirculating vortical flow in the downstream region of the cavity. This mechanism is particularly significant in the context of the present work, where our simulations indicate hydrodynamic origins of unsteady behavior on the cone-step.

Flows involving shock-boundary layer interactions (SBLIs), such as those over double-wedge/compression ramp (or its axisymmetric equivalent, the double cone) are similar to cavity flows due to impinging/reattaching shear layers. However, a major difference is the absence of an explicit cavity that anchors the separation location. Therefore, the separation is sensitive to the downstream pressure-gradient and downstream geometry.
Unsteadiness in hypersonic flows over compression ramps and double wedges has also been investigated from the perspective of global instabilities. A global instability framework was applied in the study by \citet{gs2018onset,Gs2017}, where modal characteristics of unsteadiness were examined. Earlier investigations by \citet{Robinet2007} also exist, in which global mode analyses of stable configurations were performed. % however, no unstable unsteady modes were identified in those studies.

\citet{gs2018onset}, have identified two principal types of unsteadiness in laminar shock-boundary layer interactions, each associated with distinct families of unstable modes: mode $B4$, which is linked to the three-dimensionality of the separation bubble, and mode $B5$, characterized by upstream-traveling three-dimensional waves. The labels `$B4$' and `$B5$' correspond to branches of unstable modes that emerge as the strength of the recirculating region increases. %Specifically, mode $B5$ represents a class of oscillatory three-dimensional instabilities. 
These mode families have subsequently been observed in various geometrical configurations. For instance, unsteadiness was reported in several configurations—including 0/15$^\circ$ compression ramp (\citet{Cao2021,Cao2022} and 25/55$^\circ$ double cone \citet{Hao2022}—where the B5 mode was identified as the dominant instability mechanism.

Additional investigations into instabilities in hypersonic separated flows have been undertaken by ~\citet{Tumuklu2018,Sawant2022}. These studies employed Direct Simulation Monte Carlo (DSMC) methods to examine high-Mach-number, high-Knudsen-number flows. Numerical simulations revealed the presence of Kelvin-Helmholtz (KH) instability as well as low-frequency unsteadiness that envelops the entire shock-detachment and reattachment system, which they characterize as a self-excited instability.

The discussion thus far has centered on oscillatory unsteadiness. We now turn our attention to previous works on pulsatory unsteadiness, proposed physical mechanisms, and their relevance to the present study.

\subsection{Pulsatory Unsteadiness } 

\textbf{\em Spiked axisymmetric bodies:} Spiked axisymmetric bodies are characterized by annular separation bubbles bounded by conical and bow shocks. Large unsteady motions in shock-boundary layer interactions were first reported in flows over spiked axisymmetric bodies with blunt nose tips~\citep{mair1952}. These unsteady pulsations were associated with the fluctuating motion of the oblique tip shock, which periodically merged with the blunt body bow shock. Subsequent experiments confirmed the presence of pulsatory shock oscillations in spiked cones~\citep{maull1960spike}. It was found that the cone angle and the ratio of spike length to cone length significantly influenced SBLI dynamics~\citep{wood1962}. When the cone angle exceeded the conical shock detachment angle, pulsatory oscillations were consistently observed. At lower cone angles, low-amplitude oscillations were also noted in the reflected shock and the separation zone. Similar dependencies on spike length were reported by \citet{maull1960spike} for spiked blunted cylindrical bodies at Mach 10. In further studies, \citet{holden1966spike} used a spiked-cone-cylinder geometry at Mach 10 and 15, demonstrating that small-amplitude oscillations transitioned to large-amplitude pulsations above a certain cone angle for a fixed spike length, or below a critical spike length for a fixed cone angle.

Initial interpretations of these oscillations~\citep{maull1960spike} attributed them to the high pressure in the region behind the bow shock on the blunt cylinder. However, a different mechanism was proposed by \citet{Antonov1977}, who identified a high-pressure region near the triple point, positioned between the post-bow shock zone and the near-wall flow where the oblique and bow shocks intersect. The driving mechanism for the pulsations was further explored by \citet{kenworthy1978}. A broader spectrum of axisymmetric configurations, ranging from spiked bodies to concave cones were investigated. It was concluded that the pulsations were induced by flow reversal resulting from a localized high-pressure region between the conical foreshock and the bow shock.

Further insight into the sustaining mechanisms was provided by \citet{panaras1980concave}, who identified a continuous mass influx from the free stream driven by an annular supersonic jet resulting from an Edney-4 interaction as the source sustaining the large-amplitude pulsations. However, this interpretation was challenged by detailed axisymmetric computations presented by \citet{feszty2004a} for spiked cylinders. These simulations revealed the formation of a separation vortex and demonstrated that the air inflating the separation bubble originated from within the bubble itself, where it was trapped during the early stages of each unsteady-flow cycle. Their finding pointed to an intrinsic instability mechanism capable of self-sustaining the periodic inflation and deflation observed in these high-speed flows.

\textbf{\em Planar double-wedges:} Similar pulsatory motion of the shock structure has been observed in simulations of planar shock-boundary layer interactions (SBLI) on two-dimensional double-wedges~\citep{olejniczak1996wedge}. More recently, experiments conducted at high freestream enthalpy conditions over double-wedge configurations with finite spanwise extent have revealed large-amplitude pulsations of the separation zone with increasing second-wedge angle~\citep{hashimoto2009experimental,swantek2015flowfield}. Similar to  spiked axisymmetric bodies, these pulsations are marked by the expansion of the separation zone and the subsequent formation of a bow shock over the second wedge, which periodically collapses and reforms. Under such conditions, \citet{hashimoto2009experimental} demonstrated that pulsatory unsteadiness leads to significantly elevated aerodynamic heating on the wedge surface, causing substantial thermal damage to the second wedge.

Three-dimensional numerical simulations under high enthalpy conditions using nitrogen and air have confirmed that such pulsations can develop even in spanwise homogeneous flow conditions~\citep{komives2014numerical}. The heat flux profiles obtained from these simulations showed good agreement with experimental measurements, indicating that the unsteady oscillations are predominantly planar in nature. This finding suggests that the pulsatory behavior observed in these flows arises primarily from nominally two-dimensional flow physics, even in configurations with finite spanwise extent.

To further explore the role of spanwise confinement in the geometries used in the experiments by \citet{swantek2012heat,swantek2015flowfield}, three-dimensional simulations over finite-span double wedges were conducted by \citet{Reinert2017,Reinert2020simulations}. Time-resolved numerical data revealed coherent oscillatory structures and captured the growth and collapse of the separated region associated with the pulsatory SBLI. Notably, it was found that most of the mass influx into the separation zone occurred within a narrow time window during each pulsation cycle, with time scales consistent with those reported in axisymmetric simulations of spiked bodies by \citet{feszty2004a}. The influence of vibrational non-equilibrium on the pulsation cycle was also examined, showing that while the amplitude of pulsations decreased at higher enthalpy, the frequency remained unchanged which demonstrated the robustness of the underlying fluid dynamic mechanisms responsible for sustaining the pulsatory motion across varying flow conditions.

\textbf{\em Double cones:} Pulsatory unsteadiness in axisymmetric configurations, such as spike-step and cone-step geometries has recently been studied in \cite{Hornung_Gollan_Jacobs_2021}. In these configurations, pulsatory unsteadiness was attributed to vorticity at the tip, either from viscosity at the wall or from the inviscid bow shock upstream of the cone tip. They suggested that the unsteadiness may be primarily inviscid in origin. Their work systematically explored the geometric parameter space over which the unsteady transition boundaries exist in double-cone configurations. Similarly, \cite{sasidharan2021} conducted experiments to explore a geometric parameter space and identifed the oscillatory and the pulsatory unsteadiness regimes.  More recently, \citet{Kumar_Sasidharan_Kumara_Duvvuri_2024} further analyzed the oscillations on a cone-step configuration. They proposed and calibrated an acoustic resonance model based on Rossiter's expression to predict the oscillation frequency. We will later compare this model to the current experiments in the Appendix.

\subsection{Oscillatory and pulsatory unsteadiness in inlets: Buzz instabilities}

%Similar to the axisymmetric cone step geometry, axisymmetric supersonic inlets also encounter oscillatory and pulsatory unsteadiness beyond a certain critical back pressure in the inlet. 
%\adw{R2-2: why is this relevant?} 
Similar to the axisymmetric cone step geometry, supersonic inlets encounter pulsatory unsteadiness below a certain mass flow rate in the inlet, which is typically referred to as big buzz. Unsteadiness associated with the big buzz involves the inlet shock-boundary layer interaction and is also referred to as the Dailey criterion for onset of buzz~\citep{Dailey1955}. 

Unsteadiness due to shock oscillations and flow separation in inlets has been extensively reported~\citep{Chima2012} in the past. However, only recently have simulations and experiments have begun to clarify the underlying mechanisms~\citep{Berto2020,SepahiYounsi2023}. % Unsteadiness associated with shock oscillations and separated flow in inlets has been observed for more than half a century. However, only in the past decade or two, simulations and experiments have begun to shed light on the mechanisms at work.
Prior work by \cite{Antonov1977} investigated high-speed flow over a simplified model of an inlet consisting of a needle mounted on a cylindrical cavity. Their experiments demonstrated that presence of a cavity influences the onset of buzz, but it does not affect the frequency of these pulsations, which are instead determined by the size of the flow separation region. 
\cite{trapier2007experiment, trapier2007analysis} conducted experiments and a time-frequency analysis of supersonic inlet buzz, linking low-frequency buzz to boundary layer separation and high-frequency buzz to shear layers. 
%\textcolor{red}{Interestingly, the compression ramp inlet geometry appears to have a step}. % modify the following 
Their compression ramp inlet geometry had a forward facing step.
The authors report that low-amplitude oscillations with the same frequency as large-scale buzz occur before the buzz begins and suggest that the physical mechanism responsible for buzz is already active in the flow such that the mechanism causes smaller oscillations of the normal shock at the same frequency, despite the difference in amplitude. Our results on the cone-step geometry also show similar early signs of unsteadiness before the onset of large-scale motion.

%may indicate an acoustic feedback effect. However, later studies do not support this explanation, as discussed in a later section. Our results on the cone-step geometry also show similar early signs of unsteadiness before the onset of large-scale motion.

% -- following does not flow well
% \textcolor{red}{The authors report that the low-amplitude oscillations at the same frequency as those corresponding to large scale buzz precede it. They further suggest that the physical mechanism at the origin of buzz is already present before the very beginning of the buzz, and is responsible for the smaller oscillations of the normal shock at the same frequency. They postulate that since the frequency is the same, whereas the amplitude of the shock displacement is very different, there is perhaps an acoustic feedback phenomenon. Later studies do not support this last claim as we discuss later. Similar observations will be reported in our results on the onset of unsteadiness in the cone-step geometry.}

The work of \cite{hui2011} on buzz instabilities in a Mach 5 inlet concluded that the oscillation mechanism in conventional supersonic inlets, based on acoustic resonance, does not explain hypersonic inlet buzz. Instead, they identified flow spillage at the duct as the primary source of disturbance and attributed the unsteadiness to a fluid-mechanical feedback mechanism. \cite{chima2012memo} provides a thorough summary of earlier studies by \cite{fisher1970report}, \cite{nagashima1972report}, and \cite{trapier2007experiment}, and highlights the similarities between the motion of the shock foot in inlet flows and the pulsatory shock behavior seen in cone-step configurations.

In related work, \cite{shang1980} proposed a feedback-driven mechanism to explain buzz as a self-excited oscillatory unsteadiness. They used subsonic wave propagation theory to predict the oscillation frequency and also applied Rossiter’s empirical formula by estimating the wave speeds of upstream and downstream traveling waves. Chima expanded on this by introducing the term ``Spike Buzz" to describe the pulsatory flow following boundary layer separation, consistent with findings from axisymmetric spiked flow studies. He noted that the high-frequency oscillations at the spike tip do not correspond to standing wave resonance modes but instead arise from spike buzz—a self-sustaining shock oscillation attached to a spike. Chima’s simulations further showed that the commonly used standing wave model did not apply, as the shock system responded to downstream pressure changes rather than reflecting upstream-traveling waves.

\cite{soltani2012} investigated the effect of angle of attack on buzz and found that higher angles led to increased instability and larger shock displacements. Even within high-frequency buzz regimes, both high- and low-frequency oscillations with irregular periods were observed, with low-frequency oscillations showing smaller amplitudes. As mass flux decreased, the low-frequency oscillations became dominant, indicating nonlinear amplification of an intrinsic instability. More recently, buzz in double ramp geometries was studied by \cite{Berto2020}, who concluded that the low frequencies of “big buzz” are not linked to inlet acoustic resonance. Preliminary results by \cite{grenson2018} also suggest that buzz, or shock-boundary layer interaction unsteadiness, may stem from a linear instability related to recirculation physics—similar to the unstable modal branches identified in compression ramp SBLI~\citep{gs2018onset}.

As will be shown in the subsequent analysis, our work, similar to the observations in the inlet buzz literature, also points to a self-excited oscillatory unsteadiness in the cone-step flow.  
% Tan et al. studied hypersonic buzz, noting periodic shrinking and expansion of separated regions during high-frequency buzz and bursting during low-frequency buzz.

\subsection{Motivation and layout of the paper}

\begin{figure}
    \centering
    \includegraphics[width=0.8\linewidth]{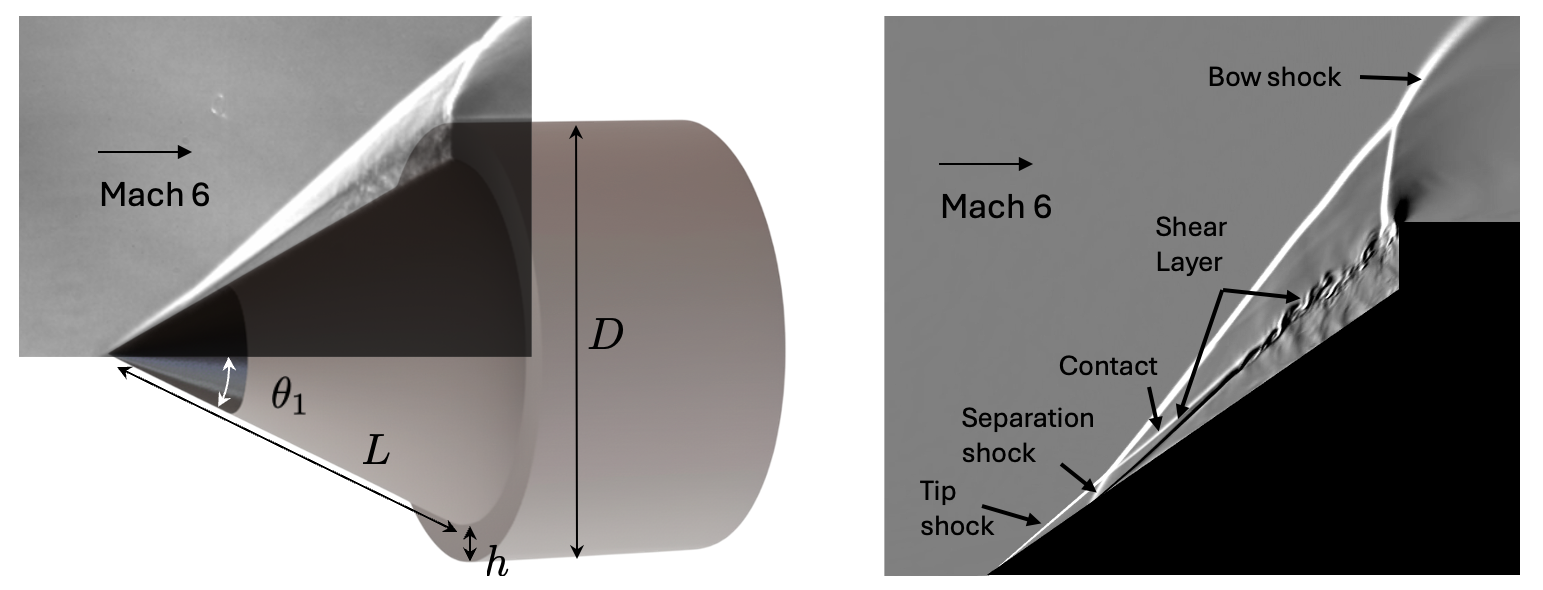}
    \caption{Flow configuration and features at Mach 6 flow on a cone-step, left: schematic of the cone-step geometry; right: flow field visualization using vertical density gradient.}
    \label{fig:schematic}
\end{figure}

In this work, we investigate the unsteadiness in the SBLI on a cone-step geometry consisting of a cylindrical forward-facing step introduced along an axisymmetric cone. {\color{blue}Figure \ref{fig:schematic} shows the key features of the shock system for this geometry.} This axisymmetric configuration avoids leading edge effects \citep{Reinert2017} and shares topological features with both double-cone geometries and spiked bodies, as well as the cone-capsule-disk configurations studied in the context of rocket abort systems~\citep{Ozawa2009,Wang2011}. 

% \begin{figure}
%     \centering
%     \fcolorbox{blue}{blue}
%     {\includegraphics[scale=0.5]{Duvvuri2021Schematic.png}}
%     \caption{Prominent flow features of the steady shock structure for cone-step geometry. Reproduced from \cite{sasidharan2021}.}
%     \label{fig:Schematic}
% \end{figure} 

Previous investigations into this geometry, using schlieren visualization and numerical simulations have documented self-sustained unsteadiness~\citep{Ozawa2009,Wang2012}. 
Exploration of the cone-step geometry and its effect on unsteadiness has been recently been carried out in the computational work of~\citet{Hornung_Gollan_Jacobs_2021} and the experimental work of ~\citep{sasidharan2021,Kumar_Sasidharan_Kumara_Duvvuri_2024}.

% As first discussed in \citet{Reinert2017}, highly pulsatory flow regimes are observed in double wedge compression ramp separation zones. It was shown in the paper, the flow is significantly affected by the role of 3D relief and mass leakage from the side walls. Therefore unsteadiness on double cones is more attractive for isolating the edge effects.

The onset of unsteadiness has previously been studied in the geometric parameter space.
We additionally consider the role of the Reynolds number Re at 
 a fixed Mach number and show that it is another important parameter that can cause the onset.
The Reynolds number allows for understanding the role of the boundary layer in the unsteadiness process. 
%Similar to how the geometric parameter space has been explored for unsteady boundaries, 
% In prior studies, the role of the Reynolds number has not been quantified and therefore will be a major focus of this work. 

An important component of the paper is role of companion simulations that allow for additional inference of the flow dynamics. A careful comparison between the experimental and the computational studies is carried out. The comparison is particularly effective due to the presence of a low-disturbance or ‘quiet’ flow environment devoid of  external disturbances that corrupt the spectral information associated with the unsteady interaction. As discussed previously, flows with separation dynamics can be significantly sensitive to the presence of external disturbances due to large amplification that occurs in the high curvature regions of the boundary layer, susceptible to centrifugal instabilities.

Our work is an important effort to investigate the unsteadiness of the shock-boundary layer interaction, in which the flow transitions from fully steady interaction, to unsteady shear layer, to mildly oscillating separation zone and finally to highly pulsating interaction.
Pulsation refers to the regime in  which the bubble moves all the way to the nose tip and then sheds. 

Different unsteadiness regimes exist in this flow due to (a) the presence of a large separation zone, which allows for the shear layer to break down prior to reattachment, and (b) the explicit presence of a reattachment/bow shock that interacts with the nose tip shock.
However, the effect of the flow parameters on the unsteadiness boundaries remains unclear and is therefore the subject of this paper.
% Spanning these regimes using a single control parameter unsteadiness regimes and transition boundaries  is not  cannot be easily are not common in coventionally studied laminar shock boundary interactions 

%The paper is organized as follows: We describe the experimental facility, the Boeing/AFOSR Mach 6 Quiet Wind Tunnel (BAM6QT). The geometric details of the cone step are discussed in . The flow conditions considered in the paper are provided. The Schlieren results in the context of the cone half angle-Re space are provided. An analysis of the unsteadiness from the Schlieren intensity measurements are carried out next. The experimental section is followed by the details on the computational setup and the CFD results. Comparison to experiments is carried out first and then the validated computational results are further analyzed for understanding the onset of the interaction. 

The paper is organized as follows. We begin by describing the experimental facility---the Boeing/AFOSR Mach 6 Quiet Wind Tunnel (BAM6QT)---used for the study. This is followed by a discussion of the geometric details of the cone-step model. The flow conditions considered in the experiments are then compared in the context of previous experimental work. Next we present schlieren visualization results as a function of the cone half-angle and the freestream Reynolds number. A spectral proper orthogonal decomposition of the observed unsteadiness is carried out. The experimental section is followed by details and results of the companion CFD simulations. Comparisons are conducted to validate the simulations and the validated CFD results are analyzed to shed light on the onset and nonlinear stages of observed flow unsteadiness on the cone-step.

\section{Experimental Methods}
\subsection{Boeing/AFOSR Mach 6 Quiet Tunnel}
\label{sec:tunnel}
% Experiments were conducted in the Boeing/AFOSR Mach-6 quiet tunnel (BAM6QT) at Purdue University. The BAM6QT is a Ludwieg tube facility with a downstream burst diaphragm system, as seen in Figure~\ref{fig:tunnel} . It has a run time of about 5 seconds. To run the tunnel, a pair of diaphragms downstream of the nozzle is burst which causes an expansion wave to propagate upstream. The expansion wave reflects between the end of the driver tube and the contraction about once every 0.2 seconds. Each reflection drops the stagnation pressure of the flow which allows a range of Reynolds numbers to be observed in a single run. The BAM6QT achieves quiet flow conditions by maintaining a laminar boundary layer on the nozzle wall. The nozzle is polished to a mirror finish to reduce roughness induced transition. The tunnel also has a bleed suction slot upstream of the nozzle throat that is connected to the vacuum tank downstream. When the tunnel is running, the slot suctions off the boundary layer at the throat, causing a new laminar boundary layer to form in its place. At the time of this experiment, quiet flow was achieved at stagnation pressures up to 225 psi.

\begin{figure}
    \centering
    \includegraphics[scale=0.3]{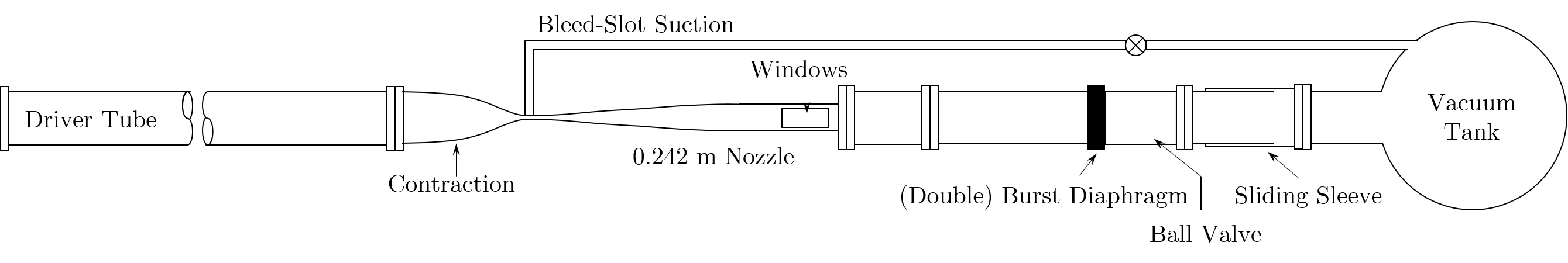}
    \caption{Schematic of the Boeing/AFOSR Mach-6 Quiet Tunnel (BAM6QT)}
    \label{fig:tunnel}
\end{figure}

{\color{blue}
Experiments were conducted in the Boeing/AFOSR Mach 6 Quiet Tunnel (BAM6QT) at Purdue University, a Ludwieg tube facility developed in the late 1990s by \cite{schneider2008}. The BAM6QT is made up of a long driver tube, a converging-diverging nozzle, a test section, a burst diaphragm system, and a vacuum tank. A schematic of the BAM6QT is shown in Figure \ref{fig:tunnel}. To run the tunnel, a pair of diaphragms are placed in the tunnel to separate the vacuum tank from the rest of the tunnel, as shown in Figure \ref{fig:tunnel}. The tunnel is then pressurized to the desired stagnation pressure, as the gap pressure between the two diaphragms is regulated to be half the difference between the two sides. The tunnel is started when the air in the gap is evacuated into the vacuum tank, causing the diaphragms to burst. An expansion wave then propagates upstream through the nozzle, and reflects between the two ends of the driver tube about every 0.2 seconds. Each reflection causes a drop in the stagnation pressure by about 1\%, resulting in a range of freestream unit Reynolds numbers observed during each run. The total run time is approximately 5 seconds before tunnel unstart occurs \citep{Mamrol2022}.

Quiet-flow operation is achieved by maintaining a laminar boundary layer on the nozzle wall, which is done through various methods in the BAM6QT. The contoured nozzle is polished to a mirror finish to minimize roughness-induced transition. The diverging section of the nozzle is elongated to reduce the growth of the G\"{o}rtler instability. A suction bleed slot just upstream of the nozzle throat is connected to the vacuum tank. This bleed slot removes the boundary layer on the nozzle wall, allowing a new laminar boundary layer to form in the diverging section of the nozzle. In quiet-flow mode, the rms pitot-pressure fluctuation level is $P_{t,\mathrm{rms}}/P_{t} < 0.01\%-0.02\%$ \citep{Mamrol2022}. During the present campaign, quiet flow was achieved at a maximum unit Reynolds number of $15.0\times10^{6}\,\mathrm{m^{-1}}$.

Freestream test conditions were reconstructed using the measured stagnation pressure and stagnation temperature in the tunnel. The initial stagnation pressure along with the stagnation pressure during the run were read from a ETL-79-HA-DC-190 Kulite pressure sensor located on the contraction section of the tunnel. A thermocouple at the upstream end of the driver tube was used to record the stagnation temperature. The tunnel is filled with air heated to a nominal temperature of 433.15 K to avoid liquefaction of the air when the static temperature drops as the air expands. The freestream static temperature is 50-60K during the steady-state portion of the run. Although it is true that the air would liquefy if it were at that temperature for a longer period of time, the flow through the nozzle at Mach 6 is too fast for this to happen on the timescale of the air's residence in the test section. The flow through the nozzle is assumed to be isentropic with a nominal Mach number of $M=6.0$ for quiet flow and $M=5.8$ for noisy flow. The static temperature and pressure are necessary to compute the unit Reynolds number, and are determined using isentropic relationships. The viscosity is found by using Sutherland’s law. A sequence of images, centered at the desired Re/m and bounded by $\pm6.41 \mathrm{ ms}$, was extracted from each shot.
}

\subsection{Model} 
\begin{figure}
    \centering    
    %\fcolorbox{blue}{blue}{
    \includegraphics[width=0.85\linewidth]{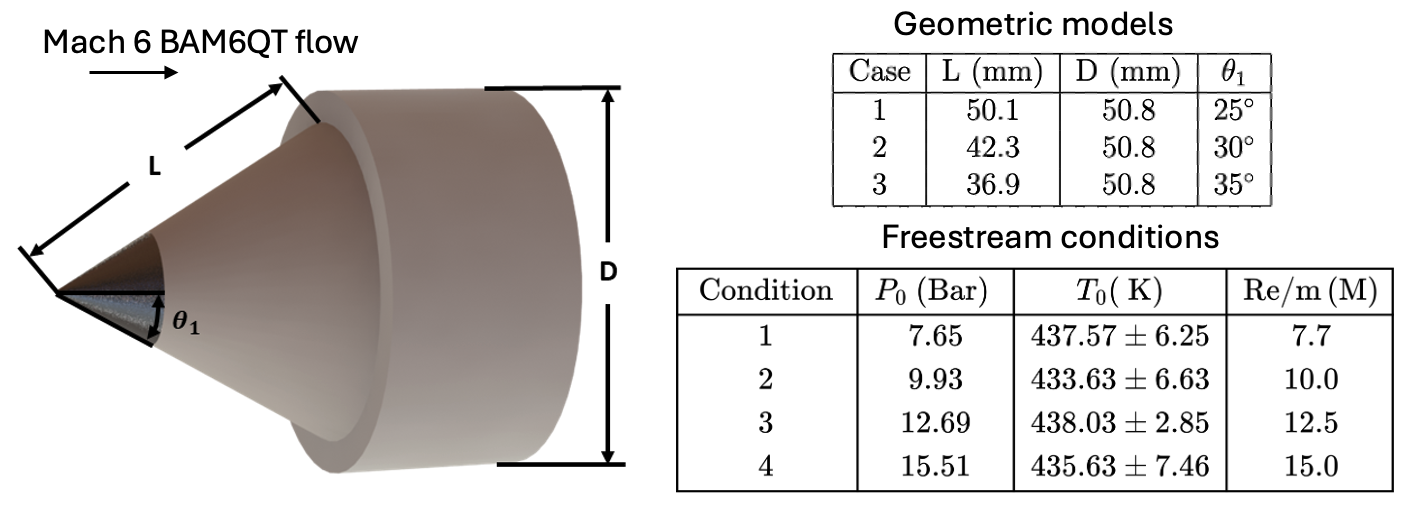}
    %}
    \caption{The cone-step geometric model and freestream conditions considered in the work.}
    \label{fig:model}
\end{figure}

The models used in this experiment consisted of a cylindrical base of diameter D with a cone of slant height L, attached to the base. Three models were constructed with the following L/D ratios and cone half-angles: $L/D=0.986$ (cone half-angle of $25^\circ$), $L/D=0.833$ (cone half-angle of $30^\circ$), and $L/D=0.726$ (cone half-angle of $35^\circ$). All three models had a base diameter D of {\color{blue}$50.8\, \mathrm{mm}$} and an axial base length of {\color{blue}$25.4\, \mathrm{mm}$}. 
The step height, i.e. the difference between the radius of the cone base and the cylinder, was constant and equal to {\color{blue}$4.3\, \mathrm{mm}$}. The geometry and run conditions are provided in Figure \ref{fig:model}.
The models were fabricated using Polyether Ether Ketone (PEEK) with an aluminum nose tip that is screwed into the frustum of the cone to protect PEEK at the test enthalpy. For the model with a cone half-angle of $35^\circ$, the metal nose tip was {\color{blue}$6.4\, \mathrm{mm}$} long. For the two models with cone half-angles of $25^\circ$ and $30^\circ$, the metal nose tip was lengthened to {\color{blue}$13\, \mathrm{mm}$} due to manufacturing constraints. Each model is internally hollow and mounted on a
streamlined sting concentric with the tunnel axis.

% Nine axisymmetric cone–step models were additively manufactured in PEEK.
% All share a cone base diameter $d=50.8$~mm; the cylindrical after–body
% diameter $D$ and cone half–angle $\theta$ were varied so that
% $D/d=\{1.2,\,1.4,\,1.6\}$ and
% $\theta=\{25^{\circ},\,30^{\circ},\,35^{\circ}\}$, giving
% $L/d=\{0.35,\,0.29,\,0.23\}$.  Threaded aluminium nose tips (0.25~in
% exposed length, extended to 0.5~in for $D/d=1.6$) protect the PEEK at the
% design enthalpy.  Each model is internally hollow and mounted on a
% streamlined sting concentric with the tunnel axis

% % AD: table here 
% \begin{tabular}{|c c c c|}
% \hline Case & L (inches) & D (inches) & $\theta_1$  \\
%  \hline 1 & 1.453& 2.000& 25$^\circ$ \\
%  2 & 1.667 & 2.000& 30$^\circ$ \\
%  3 & 1.972& 2.000& 35$^\circ$ \\
% \end{tabular}

% \begin{tabular}{|c c c c|}
% \hline Condition &  $P_0$ (Bar) & $T_0$ (K)& Re/m (M)\\ 
%  \hline  1 & 7.65  &  437.57 $\pm$ 6.25  & 7.7  \\
%          2 & 9.93  & 433.63 $\pm$ 6.63  & 10.0\\
%          3 & 12.69 & 438.03 $\pm$ 2.85  & 12.5\\
%          4 & 15.51 & 435.63 $\pm$ 7.46  & 15.0\\
% \end{tabular}

\subsection{Schlieren Setup}
Schlieren imaging was employed as the primary diagnostic tool for visualizing flow structures, leveraging the principle that light refracts when passing through mediums of varying densities. Density gradients in the flow alter the refractive index of the air, which is related to density by the Gladstone-Dale relation~\citep{settles2001}. 

We implemented a Z-type schlieren arrangement, constrained by the available optical bench size. The light path began with a CAVILUX laser source~\citep{CavitarOEM} connected via fiber optic cable to an iris on an optical bench. Passing through the iris created a controllable point source whose brightness could be adjusted using the size of the iris' opening. A four-inch flat mirror redirected the diverging light onto an eight-inch parabolic mirror, which collimated the beam. This parallel light traversed the wind tunnel test section windows and struck a second eight-inch parabolic mirror. This mirror refocused the light, which a second four-inch flat mirror then directed toward the knife-edge. The knife-edge, positioned at the focal point, selectively blocked refracted light rays and thereby visualized density gradients as variations in image intensity. To minimize optical aberrations, the angles between incoming and outgoing light paths at the parabolic mirrors were kept small and configured to be equal in magnitude but opposite in direction, to help mitigate coma. We reduced the effects of astigmatism by using the knife edge and mirrors with long focal lengths.

A Phantom TMX 7510 high-speed camera~\citep{PhantomTMXDatasheet} captured the image sequences using a frame rate of $78,000$ FPS and a resolution of $768 \times 768$ pixels. This combination provided a sufficiently large field of view to encompass the entire shock structure ahead of the model while maintaining a temporal resolution orders of magnitude higher than the expected frequencies of shock oscillations.  A convex lens was positioned between the knife-edge and the camera to adjust the focus and magnify the desired portion of the image onto the camera sensor.

%The primary diagnostic used for this study is schlieren imaging, a flow visualization technique that has been used for decades to study shockwaves and expansion fans in supersonic flows \citep{settles2001}. An overhead schematic of the setup is shown in Figure~\ref{fig:SchlierenSetup}. The two four-inch flat mirrors are needed as the BAM6QT optical bench is not long enough for the focal length of the eight-inch parabolic mirrors. A CAVILUX Smart pulsed diode laser was used as the light source for the schlieren images taken. It can pulse up to a maximum frequency of 100,000 Hz, which was sufficient for this experiment. High-speed video of the flow was recorded using a Phantom TMX 7510 high speed camera. The maximum framerate of the camera is 1.75 MHz with a maximum resolution of 1280 x 800 pixels. For this experiment, a framerate of 76,000 and a resolution of 1280$\times$800 pixels was used.  
% \begin{figure}
%     \centering
%     \includegraphics[scale=0.5]{Chase and Eric/Schlieren Setup.png}
%     \caption{Overhead view of the schlieren setup}
%     \label{fig:SchlierenSetup}
% \end{figure} 

\section{Quiet Wind tunnel experiments of cone-step flow: Effect of freestream Reynolds Number and cone half angle $\theta_1$}

\subsection{Unsteady Boundaries in the geometry and flow parameter space}

In the introduction, we outlined the relevant literature, with emphasis on unsteady boundaries for \textbf{spiked axisymmetric bodies}. These boundaries serve as a reference for interpreting the present results in the broader context of prior studies. 

%\adw{R2-4: laminar boundaries assumed for all Re} 
Motivated by previous work, we employ two geometry parameters to systematically explore the onset of unsteadiness in the geometric parameter space. These are: cone length-to-diameter ratio ($L/D$) and the cone half angle $\theta_1$. 
{\color{blue}In addition to the geometric parameters, we consider the effect of Reynolds number, as the flow parameter. In the experiments, the free-stream Reynolds number per unit meter Re$_\infty$/m is varied.  
Throughout this paper, we predominantly use $Re_L$ as the non-dimensional parameter to characterize the cases studied.
Here, $Re_L$ is associated with the boundary layer Reynolds number. We also specify $Re_D$, associated with the blunt body frontal area Reynolds number. 
Here \begin{equation}
    \text{Re}_{L} = \frac{\rho_\infty U_\infty L}{\mu_\infty (T_\infty)}, \ \     \text{Re}_{D} = \frac{\rho_\infty U_\infty D}{\mu_\infty (T_\infty)}
\end{equation}
Table \ref{tab:relred} shows the Reynolds numbers considered in this study.

\begin{table}
    \centering
    \begin{tabular}{c c c c c}
$\theta$ -- Re/m & 7.7M & 10M & 12.5M & 15M \\ \hline
% $25^\circ$ & $3.9$ & $5.0$ & $6.3$ & $7.5$ \\
% $30^\circ$ & $3.3$ & $4.2$ & $5.3$ & $6.3$ \\
% $35^\circ$ & $2.9$ & $3.7$ & $4.6$ & $5.5$ \\   
$25^\circ$ & $3.9/3.9$ & $5.0/5.1$ & $6.3/6.4$ & $7.5/7.6$ \\
$30^\circ$ & $3.3/3.9$ & $4.2/5.1$ & $5.3/6.4$ & $6.3/7.6$ \\
$35^\circ$ & $2.9/3.9$ & $3.7/5.1$ & $4.6/6.4$ & $5.5/7.6$ \\
    % $25^\circ$ & $3.9 $ & $5.0 \times 10^5$ & $6.3 \times 10^5$  & $7.5 \times 10^5$  \\
    % $30^\circ$ & $3.3 \times 10^5$ &  $4.2 \times 10^5$  &  $5.3 \times 10^5$  & $6.3 \times 10^5$ \\
    % $35^\circ$ & $2.9 \times 10^5$ &  $3.7 \times 10^5$  &  $4.6 \times 10^5$ & $5.5 \times 10^5$\\
    \end{tabular}
    \caption{Re$_L$/Re$_D$ ($\times 10^5$) corresponding to the case matrix in the experiments.}
    \label{tab:relred}
\end{table}}
%For clarity, Re$_\infty$/m is hereafter referred to simply as the Reynolds number (Re).

%Henceforth, the paper will refer to Re$_\infty$/m as the Reynolds number (Re) for convenience to the reader.

\begin{figure}
    \centering
    \begin{overpic}[width=1\linewidth]{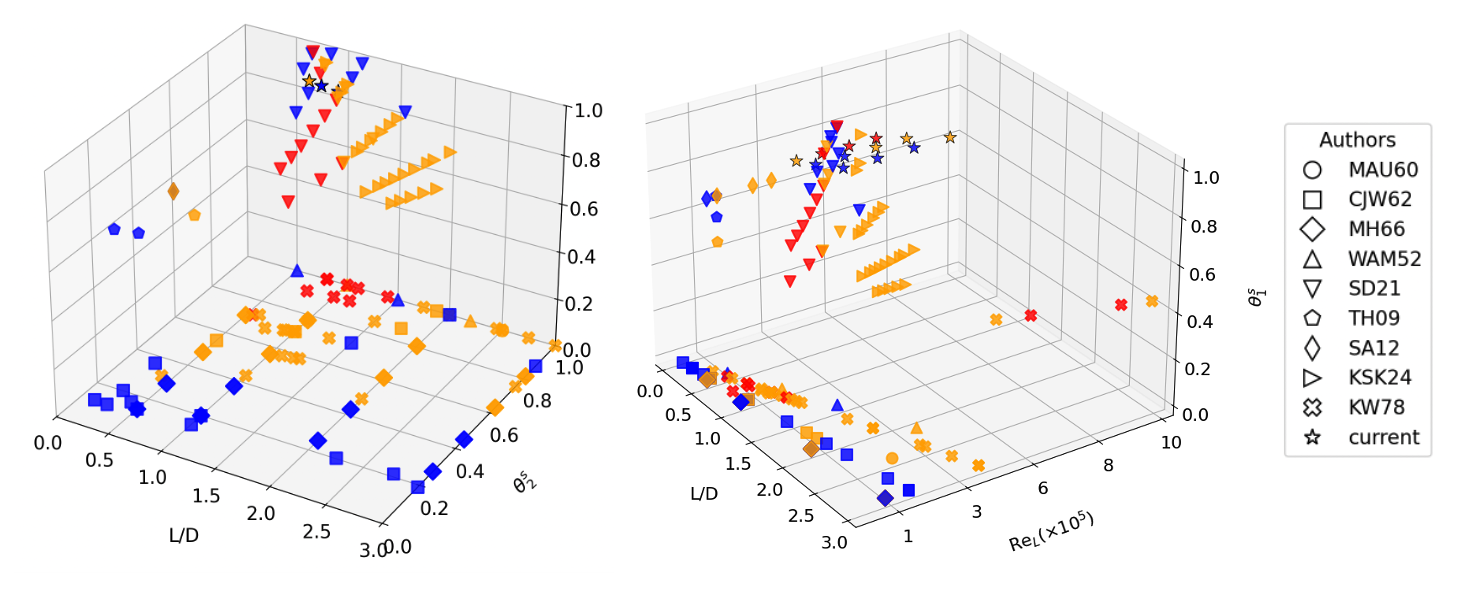}
        \put(4,38){(a)}  % (x%, y%)
        \put(50,38){(b)}  % (x%, y%)
    \end{overpic}
    \caption{An overview of the experiments in geometric and freestream Reynolds number parameter space. $\theta^s_1$ and  $\theta^s_2$ are the scaled angles defined in Equation \eqref{eq:thetadef}. The color key is blue: steady, orange: oscillatory and red: pulsatory.  The key for the references is: MAU60: \citep{maull1960spike}, CJW62: \citep{wood1962}, MH66: \citep{holden1966spike}, WAM52: \citep{mair1952}, SD21: \citep{sasidharan2021}, TH09: \citep{hashimoto2009experimental}, SA12: \citep{swantek2012heat}, KSK24: \citep{Kumar_Sasidharan_Kumara_Duvvuri_2024}, KW78: \citep{kenworthy1978}
    %\textcolor{red}{The key for the references is \ldots}}
    }
    \label{fig:exp_para_span}
\end{figure}

In figure \ref{fig:exp_para_span} (a), we have collated the available experimental data  within this three-dimensional parameter space. To facilitate interpretation, we rescale the cone half angle variables as 
\begin{align}
\label{eq:thetadef}
    \theta^s_{1} &= \frac{\theta_1 }{\theta_m} ,\quad \theta_m \equiv \tan^{-1}\left(\frac{D}{2L \cos \theta_1}\right) \\
    \theta^s_{2} &= \frac{\theta_2 - \theta_m}{ \pi/2 - \theta_m}.
\end{align}
% This ensures $\theta^s_1$ can be sampled independently of $L/D$. The decoupling allows for easier interpretation of the parameter space, in the context of the unsteadiness boundaries. 
This ensures that \( \theta^s_1, \theta^s_2 \in (0, 1) \).
Here, $\theta_1$ is the half-angle of the first cone and $\theta_2$ is the half-angle of the second cone. For a cone-step, $\theta_2=90^\circ$ and is not a variable in this study.

The experimental data from the current study is represented by \textbf{star markers} in the figures. Figure~\ref{fig:exp_para_span} (a) \upd{ shows the distribution of these data points in $L/D$, $\theta^s_2$, and $\theta^s_1$ space}. The influence of all three parameters is evident in the spread. Similarly, Figure~\ref{fig:exp_para_span} (b) presents the same dataset in the Re$_L$ space. The bottom plane of this plot represents spiked cone geometries, which exhibit a distinct demarcation at approximately $\theta^s_2 \approx 0.5$, beyond which the flow exhibits unsteadiness. The color scheme used throughout is as follows: \textbf{blue denotes steady flow, orange indicates oscillatory unsteady behavior, and red corresponds to pulsatory unsteadiness.}

\begin{figure}
    \centering
    \includegraphics[width=0.75\linewidth]{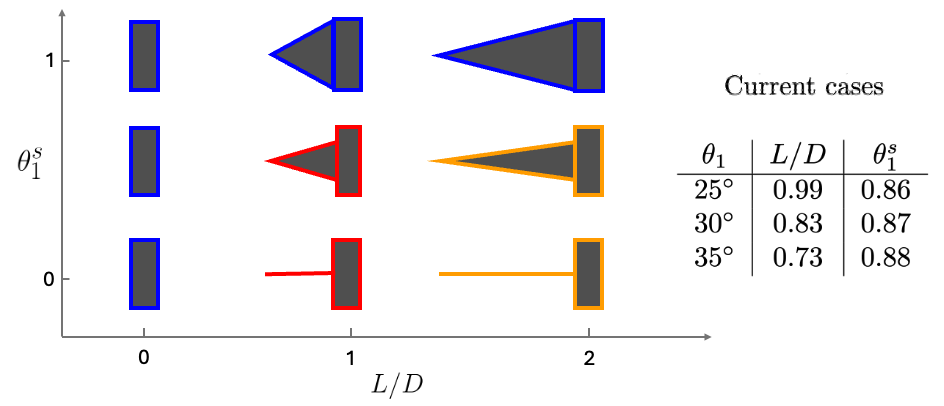}
    \caption{Geometric illustration of the parameter space $\theta^{s}_1$ and $L/D$. The color key is blue: steady, orange: oscillatory and red: pulsatory. The color coding in the illustration helps interpret collated experimental data in the $L/D-\theta^{s}_1$ space.}
    \label{fig:para_space}
\end{figure}

We now orient the reader to the theoretically expected locations of unsteady boundaries in this parameter space in figure \ref{fig:para_space}.  The scaled angle $\theta_{1s}$ facilitates a less skewed mapping of the allowable geometric parameter space. The limiting case of $L/D = 0$ and $\theta_{1}^s = 1$ corresponds to cylindrical and conical geometries, respectively. They are both associated with steady flow behavior. In contrast, configurations with $L/D > 0$ and $\theta_{1}^s < 1$ exhibit unsteady dynamics. Spiked geometries tend toward $\theta_{1}^s \approx 0$, while cone-step configurations, which is the primary focus of the present study occupy intermediate $\theta_{1}^s$ values. For cone-step geometries, when the aspect ratio is small (i.e., lower $L/D$), the flow resembles that around blunt bodies, characterized by large-scale separation and a pulsatory nature. On the other hand, larger $L/D$ values lead to thicker boundary layers. These undergo separation sufficiently downstream of the nose tip and form high-speed separation zones of the compression-ramp type which support oscillatory behavior. 

\begin{figure}
    \centering
    \includegraphics[width=1\linewidth]{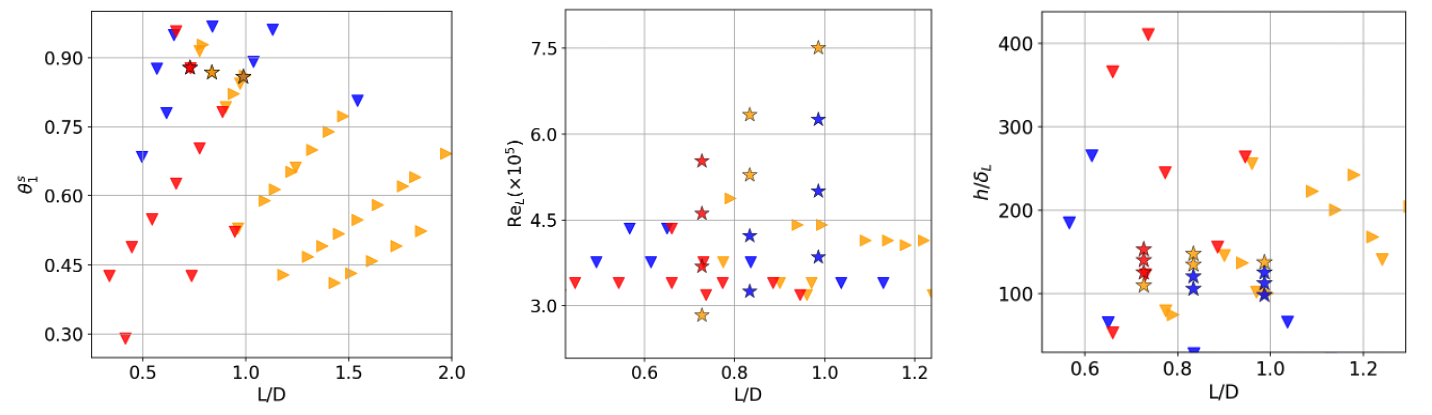}
    \caption{ An overview of the cone-step $\theta_2 = 90^\circ$ experiments in geometry and freestream Reynolds number parameter space. The figure (c) plots the non-dimensional inverse boundary layer thickness. 
    %\sidgs{R1-8 The boundary layer scaling variable helps demarcate the unsteadiness boundary for the data from quiet-tunnel experiments (plotted as stars).}}%\adw{blue}{make this into 2 figures; keep 4a and b together 
    }
    \label{fig:cone_step_exp}
\end{figure}

Following the illustration, we plot the collated data in the two-dimensional $\theta^{s}_1-L/D$ parameter space, as shown in Figure~\ref{fig:cone_step_exp} (a). The experimental data can be clearly seen to follow the unsteadiness boundaries discussed previously. 
%Figure~\ref{fig:cone_step_exp} illustrates the expected boundaries for unsteady behavior across the cone-step geometry configurations. 
Symbols corresponding to the experimental cases are explained in the caption of Figure~\ref{fig:exp_para_span}. The data from the current experiments sweep nearly horizontally ($\theta^s_1 \approx \text{const.}$) in the $\theta^{s}_1-L/D$ plane, spanning the steady$-$oscillatory$-$pulsatory regimes \comment{a critical transition region where both the steady$-$oscillatory and oscillatory$-$pulsatory transition boundaries are observed}. {As seen in figure \ref{fig:cone_step_exp}(b), an important parameter varied in our experiments is the freestream unit Reynolds number Re/m and equivalently the Reynolds number Re$_L$ (and Re$_D$).}
This is a \textbf{unique exploration} of the parameter space carried out in the present work. 
%\textcolor{red}{Re/m is shown to have a similar influence on the flow unsteadiness, as the geometric parameter $\theta_1$ or $D/L$.}
%To rationalize this observation, we propose 
%the flow parameter of relevance that explains this similarity i.e. the role of Re/m and $\frac{1}{L/D}$ in controlling the onset of unsteadiness. Specifically, we compute the dependence of the boundary layer thickness $\delta$ on the cone at the separation location on the geometric parameters. 
As is the case with decreasing $L/D$, we hypothesize increased \textcolor{blue}{freestream Reynolds number} leads to a thinner boundary layer and is more likely to separate than a thicker boundary layer. Therefore, the non-dimensional cone boundary layer thickness prior to separation is as an important flow parameter for characterizing the state of unsteadiness of the flow. {\color{blue}The relevant non-dimensional length scale for the separating boundary layer is the step height $h$ (marked in Figure \ref{fig:model}).
This is corroborated by the fact that the separation location does not change significantly with change in $\theta_1$ for the range of cone step parameters considered. Then inverse non-dimensional cone boundary layer thickness scales as
\begin{equation}
    \frac{h}{\delta_L} = \left(\frac{1}{2(L/D)} - \sin \theta_1 \right)\sqrt{\text{Re}_{L,e}}
\end{equation}
where \begin{equation}
    \text{Re}_{L,e} =  \frac{\rho_e U_e L}{\mu_e} = \text{Re}_{L,e} (\rho_\infty, U_\infty, T_\infty,\theta_1)
\end{equation}}
Here, the dependence on $\theta_1$ arises from the two-dimensional laminar boundary layer thickness in $\xi-\eta$ co-ordinates which is related to a laminar boundary layer
on a body of revolution using the Mangler transformation \citep{mangler1948zusammenhang, tetervin1969generalization} given by 
\begin{align}
\xi(x) = \frac{1}{L^2} \int_0^x r^2(x) \, dx, \quad 
\eta(x, y) = \frac{r(x)}{L} \, y, \quad r(x) = x \sin \theta_1
\end{align}
Here $\xi,\eta$ are the planar boundary layer co-ordinates and $r(x)$ is the sectional radius of the cone.
% Use Re x formula and R = L sin theta to get the 1/delta expression. Don't overcomplicate. 
 \upd{ The boundary layer edge Reynolds number Re$_L,e$ is computed using the inviscid post-conical shock state ($\rho_e,U_e,T_e,\mu(T_e))$, which is a function of the cone angle $\theta_1$ and is obtained via the inviscid Taylor-Maccoll solution. }
As can be seen in figure \ref{fig:cone_step_exp}(c), the inverse non-dimensional boundary layer thickness $h/\delta_L$ correlates with onset of unsteadiness for a given $L/D$. We note that this variable has an implicit dependence on $L/D$.
% or freestream \textcolor{blue}{Re$_L$} is increased.
% So, our experimental data, a 2D grid in the $L/D-$ \textcolor{blue}{Re$_L$} space is effectively a 1D variation over the boundary layer thickness via the parameter $\sin \theta_1 \sqrt{\text{Re}_e}$.
As will be discussed later, the origin of unsteadiness is a complex process involving the interaction of different units in the flow-field. However, $h/\delta_L$ provides a parsimonious parameter for engineering predictions of the separation unsteadiness.
% \textcolor{red}{Therefore, the role played by the boundary layer development upstream of separation. }
The combined geometric and viscous boundary layer effects together determine the precise onset of unsteady behavior in the flow.

\subsection{Onset and Progression of Unsteadiness: \\ Influence of Reynolds Number and Geometry}
\begin{figure}
    \centering
    \includegraphics[width=1\linewidth]{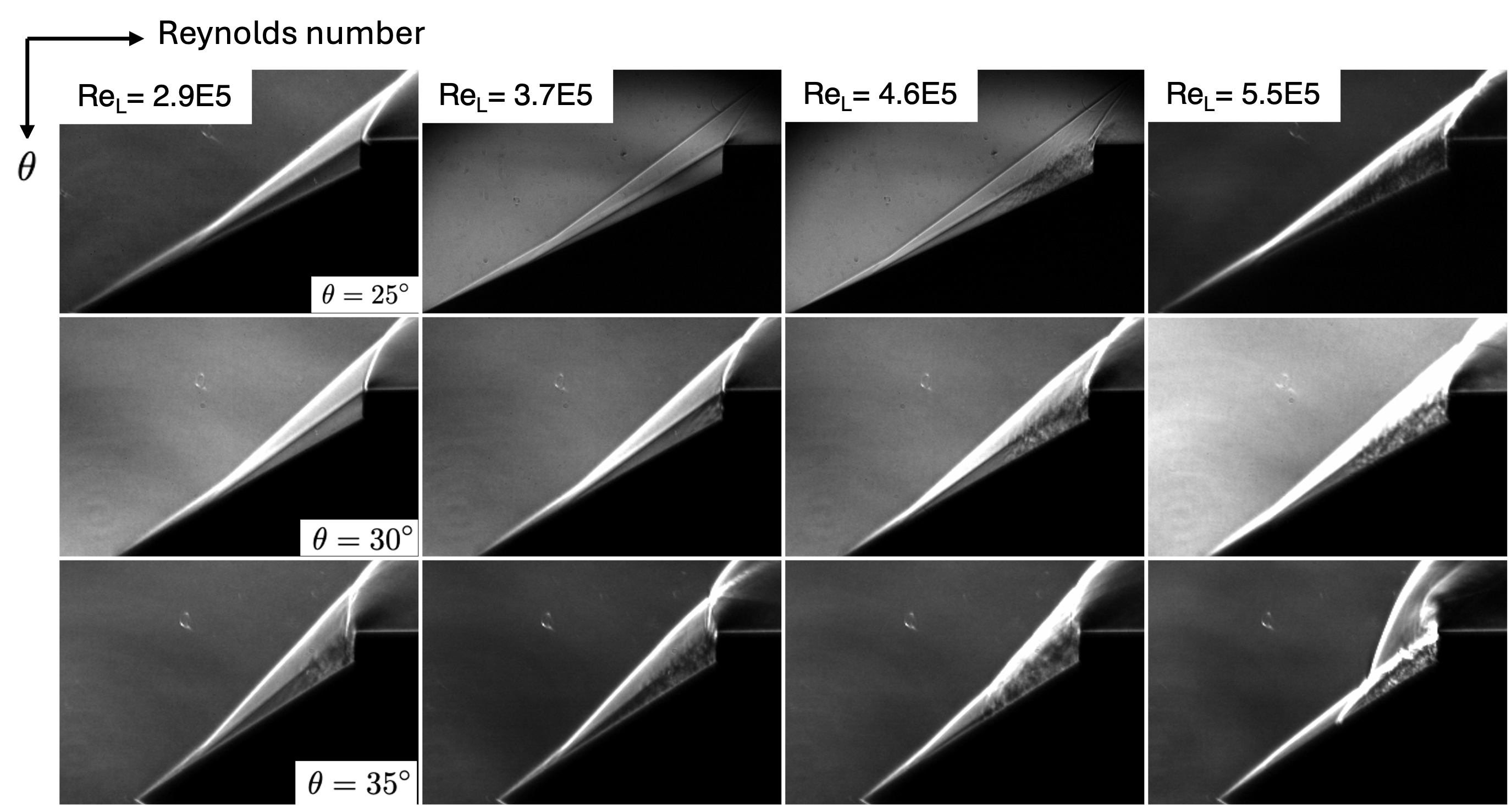}
   \caption{Experimental schlieren frames organized by the Re$-\theta$ parameter space explored in the current work. Our work traverses the unsteadiness boundary in both parameters.}
    \label{fig:exp_para_schl}
\end{figure}

This section details the unsteady flow phenomena observed in the experiments. As shown in Figure~\ref{fig:exp_para_schl}, a strong dependence of flow unsteadiness on both the cone angle and the \textcolor{blue}{Reynolds number Re$_{L}$} is observed. Increasing the cone angle induces a thinner boundary layer which separates slightly early. The shear layer is unstable to unsteady fluctuations and breaks down at a location where it comes in proximity to the contact emanating from the shock-shock interaction. This will be discussed in detail later in Section \ref{sec:lst_contact}. For a fixed geometry, increasing the \textcolor{blue}{Reynolds number} has a similar effect on the boundary layer separation leading to the onset of unsteadiness. The progression from steady to highly unsteady flow, as either \textcolor{blue}{Re$_L$} or cone angle is increased, follows a sequence broadly consistent with observations in SBLI literature. The sequence has the following three phases: 
\begin{enumerate}
    \item \textbf{Shear Layer Instability:} Unsteady fluctuations first occur within the free shear layer separating from the cone surface. The separation shear layer is unstable to 3D Kelvin-Helmholtz (KH) instabilities \cite{}. These lead to fluctuations in the shear layer's position and structure, resulting in an unsteady reattachment process near the corner/step.
    
    \item \textbf{Separation Point Oscillation:} With a further increase in \textcolor{blue}{Re$_L$} or cone angle, the separation point itself begins to exhibit noticeable axial oscillations. The unsteadiness is no longer confined to the downstream part of the separated region but affects the global structure of the separation bubble.
    \item \textbf{Pulsating Motion:} At increased \textcolor{blue}{Re$_L$} or cone angles, the oscillations enters a highly nonlinear regime characterized by large-scale, low-frequency oscillations often termed 'pulsation'. In this regime, the separation point undergoes significant upstream and downstream excursions, moving cyclically towards the cone's nose tip. This large-scale breathing motion of the separation bubble is strongly coupled with the dynamics of the bow shock formed ahead of the step/cylinder. In extreme phases of the cycle, the bow shock momentarily detaches from the cone and later convects downstream before reforming. This pulsation represents a fully developed, highly nonlinear state stemming from the oscillations described in the second  phase of the sequence.
\end{enumerate}

\textcolor{blue}{The current experiments emphasize the critical role of the Reynolds number Re$_L$} in defining the boundary between steady and unsteady cone-step SBLI, complementing the  influence of geometric parameters and Mach number discussed in prior studies. Previous characterizations of unsteadiness boundaries in the literature primarily emphasize Mach number-geometry space, often operating under the assumption of sufficiently high \textcolor{blue}{Re$_L$}. Our findings underscore the critical role of \textcolor{blue}{Re$_L$ as an additional parameter} governing the onset and nature of unsteadiness at the transition boundary.

\subsection{SPOD Analysis of schlieren intensity images}
\label{sec:exp_spod}
% \rev{R3-1: is Re7.7 oscill or steady?}
% \rev{R3-2: SPOD settings and the convergence}
To gain quantitative insights into the dominant frequencies and spatial structures of the associated unsteady behavior,  Spectral Proper Orthogonal Decomposition (SPOD) was performed on the time-resolved experimental schlieren  data~\citep{Lumey2012,Schmidt2020}. The $35$-degree cone configuration was selected for this detailed analysis as it spans the range from  \upd{nearly steady}  separation at lower \textcolor{blue}{Re$_L$} to the pronounced pulsating mode at the highest \textcolor{blue}{Re$_L$}. \upd{The details of the SPOD analysis are summarized in the Appendix \ref{spod_details}.}
% \sidgs{R1-9: Details of the SPOD}

\begin{figure}
    \centering
\includegraphics[width=1\linewidth]{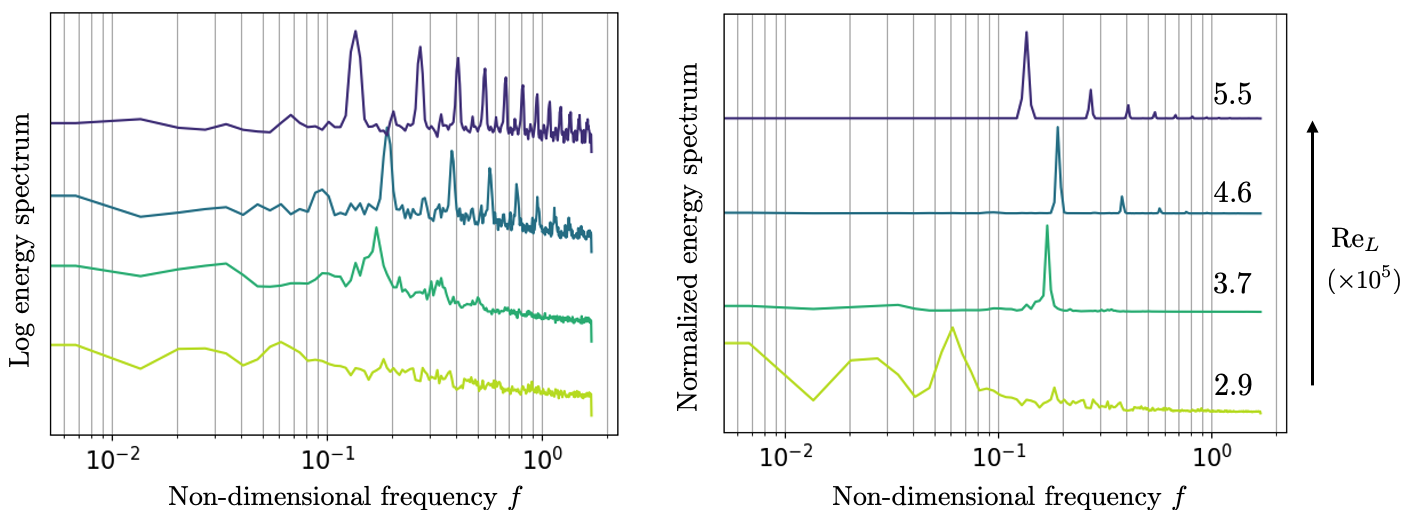}
    \caption{SPOD spectrum from experimental schlieren intensity images with \textcolor{blue}{Re$_L$} at $\theta = 35 ^\circ$. The peak for \textcolor{blue}{Re$_L$}=3.7,4.6 $\times10^5$ is at St $\approx0.17$ and for \textcolor{blue}{Re$_L$}=5.5 $\times10^5$ at St $\approx0.13$.
    Please note that the spectra for each case is arbitrarily shifted on the y-axis.}
    \label{fig:SPOD_exp_spect}
\end{figure}

The SPOD analysis identifies the characteristic frequencies of the coherent structures within the flow is shown in Figure~\ref{fig:SPOD_exp_spect}. The non-dimensional frequency is expressed as the Strouhal number, $f = f^* L / U_\infty$, where $f^*$ is the dimensional frequency, $L$ is the characteristic length scale which we take to be cone's slant height from nose-tip to base, and $U_\infty$ is the freestream velocity.  %\adw{R2-1: comment on higher values when compared with turb SBLI oscillations} 
At the onset of significant large-scale unsteadiness, observed in the separation point oscillation, the dominant mode corresponds to a Strouhal number St = $f \approx 0.17$, which falls in the range 0.15-0.25, typical of unsteady frequency oscillations observed in spiked axisymmetric bodies at supersonic and hypersonic speeds. This dominant Strouhal number remains approximately constant as the \textcolor{blue}{Reynolds number} is increased up to \textcolor{blue}{Re$_L$} = 4.6 $\times10^5$ case, even as the amplitude of the unsteadiness grows. However, at higher \textcolor{blue}{Reynolds number Re$_L$} = 5.5 $\times10^5$, which exhibits the fully developed, large-amplitude pulsating motion, the dominant frequency drops slightly, yielding St $\approx 0.13$. This decrease in frequency for the most energetic mode is a common characteristic observed in fluid systems entering highly nonlinear regimes in post-bifurcation dynamics, often associated with the establishment of large amplitude limit-cycle oscillations as discussed in Section~\ref{sec:SSM}

    \begin{figure}
        \centering
        \includegraphics[width=1\linewidth]{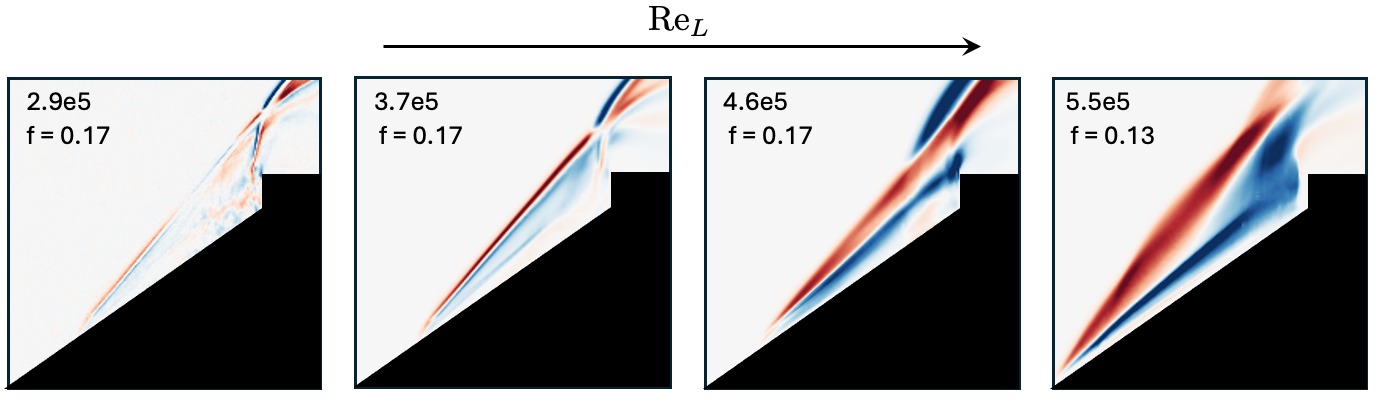}
        \caption{Spatial structure of the leading SPOD mode from experimental schlieren images with \textcolor{blue}{Re$_L$} at $\theta = 35 ^\circ$. The mode is representative of fluctuating intensity.}
        \label{fig:SPOD_exp_opt_mode}
    \end{figure}

Schlieren intensity maps associated with the dominant modes $(St \approx 0.13 - 0.17)$ can be utilized to interpret the underlying low frequency unsteadiness mechanism as shown in Figure~\ref{fig:SPOD_exp_opt_mode}. At moderate \textcolor{blue}{Re$_L$} when the amplitude of unsteadiness is small, the dominant SPOD mode clearly captures a coupled oscillation involving the separation shock and the bow shock formed at the step/corner. The mode shows that the oscillations in the separation shock are correlated with the oscillations in the bow shock. More importantly, this low-frequency mode is distinct from higher-frequency structures associated with Kelvin-Helmholtz instabilities within the shear layer. This is because the spatial signature of the St $\approx 0.17$ mode does not prominently feature structures localized in the shear layer.

%The spatial signature of the $St \approx 0.17$ mode does not prominently feature structures localized solely to the shear layer flapping. 
Therefore, the present analysis points to a hydrodynamic coupling of the separation zone (shock-shear layer-bubble) and bow shock as the mechanism associated with the dominant low-frequency mode. 
The reader is directed to the Appendix for more details.

\begin{figure}
    \centering
    \includegraphics[width=1\linewidth]{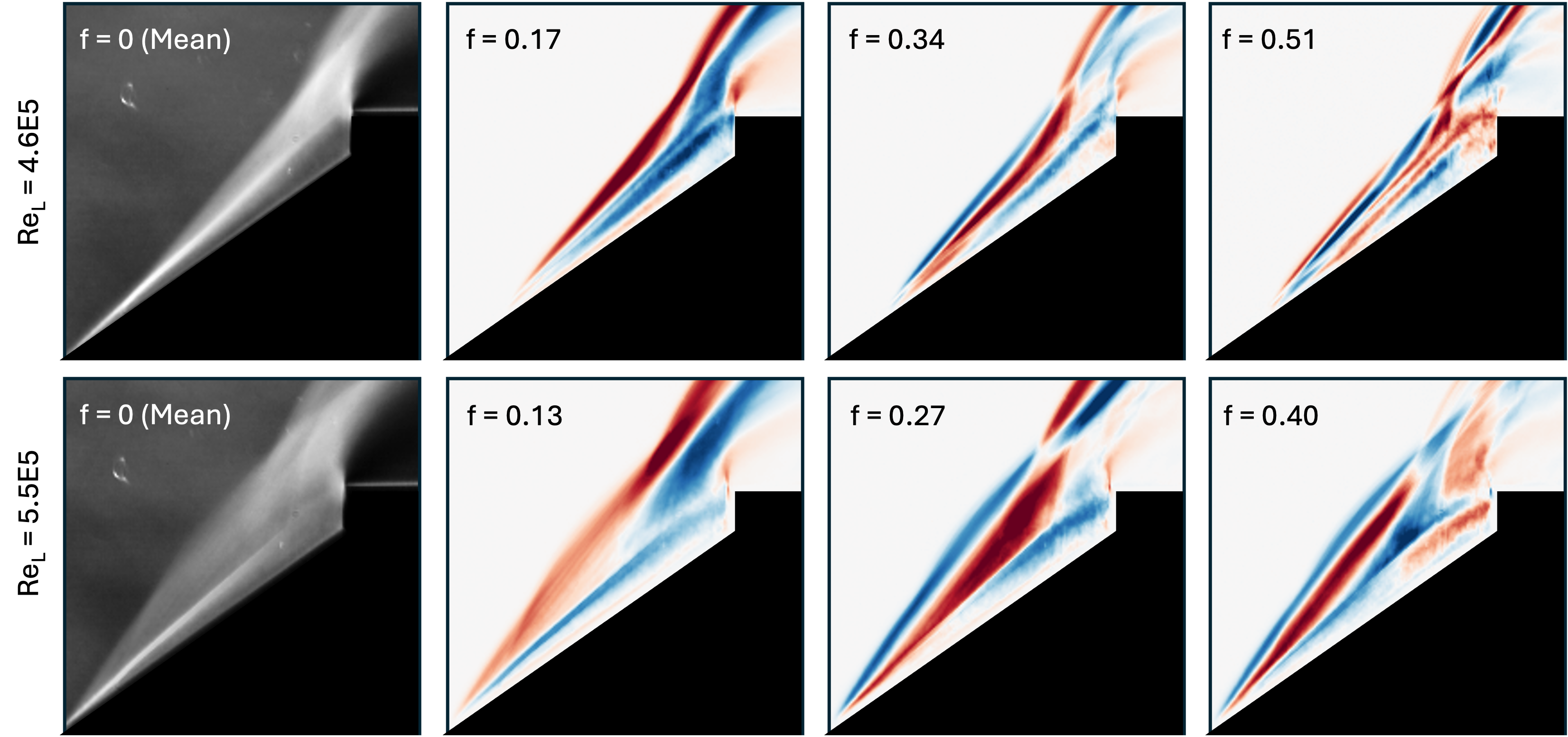}
    \caption{Spatial structure of the mean, the leading SPOD mode and its harmonics from experimental schlieren images with \textcolor{blue}{Re$_L$} at $\theta = 35 ^\circ$. Here, $f = f^*L/U_\infty$ is the non-dimensional frequency.}
    \label{fig:SPOD_exp_subh_mode}
\end{figure}

Figure~\ref{fig:SPOD_exp_subh_mode} shows the SPOD modes corresponding to the harmonics for \textcolor{blue}{Re$_L$} = 4.6$\times10^5$ and \textcolor{blue}{Re$_L$} = 5.5$\times10^5$ cases (presented in Figure~\ref{fig:SPOD_exp_spect}). 
%We do not observe Kelvin-Helmholtz structures or \sidgs{ R1-10 acoustic wave patterns in these modes as well}. 
% Their absence does not negate the importance of the shear layer breakdown which can modulate the unsteadiness of SBLI system. Unlike typical supersonic wind tunnels, this study is conducted in a quiet facility which helps isolate the shear layer dynamics from receptivity to tunnel-induced disturbances. 

 At the highest \textcolor{blue}{Reynolds number} (\textcolor{blue}{Re$_L$} = 5.5$\times10^5$ case), the oscillatory dynamics is significantly more nonlinear and the global unsteadiness spans the entire length of the cone. The corresponding SPOD mode reflects this increased coupling between the separation zone and the bow shock, both of which undergo large, correlated excursions during the pulsation cycle.

% Previous characterizations of unsteadiness boundaries in the literature focus on Mach number-geometry space, sometimes implicitly assuming a sufficiently high \textcolor{red}{Re} or neglecting its quantitative influence. Our findings highlight the importance of \textcolor{red}{Re} as an independent parameter governing the onset and nature of unsteadiness at the transition boundary. 

Next, we present the companion computational fluid dynamics (CFD) simulations of the quiet tunnel experiments.

\section{High-fidelity numerical simulations}
\subsection{Governing equations}
We solve the unsteady compressible Navier–Stokes equations in conservative form:
\begin{align}
\label{eq:nse}
\dfrac{\partial \b{U}}{\partial t} 
\,+\,
\dfrac{\partial \b{F}_j}{\partial x_j}
\,+\,
\dfrac{\partial \b{F}_j^v}{\partial x_j}
\;=\;
0
\end{align}
where $\b{U}$ is the vector of conserved solution variables, $\b{F}_j = \b{F}_j(U)$ is the inviscid flux vector and $\b{F}^v_j $ is the viscous flux vector.
\begin{align}
    \b{U}
    \;=\;
    \left(\begin{array}{c}
    \rho \\
    \rho u \\
    \rho v \\
    \rho w \\
    E
\end{array}\right),
\,
\b{F}_j
\;=\;
\left(\begin{array}{c}
\rho u_j \\
\rho u u_j+p \delta_{1 j} \\
\rho v u_j+p \delta_{2 j} \\
\rho w u_j+p \delta_{3 j} \\
(E+p) u_j
\end{array}\right),
\,
\b{F}^{v}_j
\;=\;
\left(\begin{array}{c}
0 \\
\sigma_{1 j} \\
\sigma_{2 j} \\
\sigma_{3 j} \\
\sigma_{i j} u_i+q_j
\end{array}\right)
\end{align}
where $\rho$ is density, $u_j$ is the $j^{th}$ velocity component, $E = E_{\rm int} + \frac{1}{2} \rho u_k u_k$ is total energy, and $E_{\rm int}$ is internal energy of the gas. The viscous momentum flux is $\sigma_{i j}=-\mu\left(S_{i j}-2 / 3 S_{k k} \delta_{i j}\right)$, and the molecular heat diffusion is $q_i=-\kappa \partial_j T$. $S_{ij}$ is the strain rate tensor. The viscosity of air is computed using Sutherland's law of viscosit $\mu = \mu(T)$. Conductivity $\kappa$ for air is evaluated corresponding to a Prandtl number $Pr = 0.72$. 

% \subsection{Compressible Navier-Stokes equations}

% \subsubsection{Computational domain}
% The computational domain for the simulation is shown in Figure~\ref{fig:cfd_domain}. \adw{red}{ Both axisymmetric and three dimensional computations are carried out to \ldots }

% - 2D vs 3D
% - Azimuthal domain
% - Resolution
% - Grid and domain independence
% - Time domain: Development time and statistics for the cycles. 

\subsubsection{Computational Domain and Simulation Setup}
\begin{figure}
    \centering
    \includegraphics[width=0.85\linewidth]{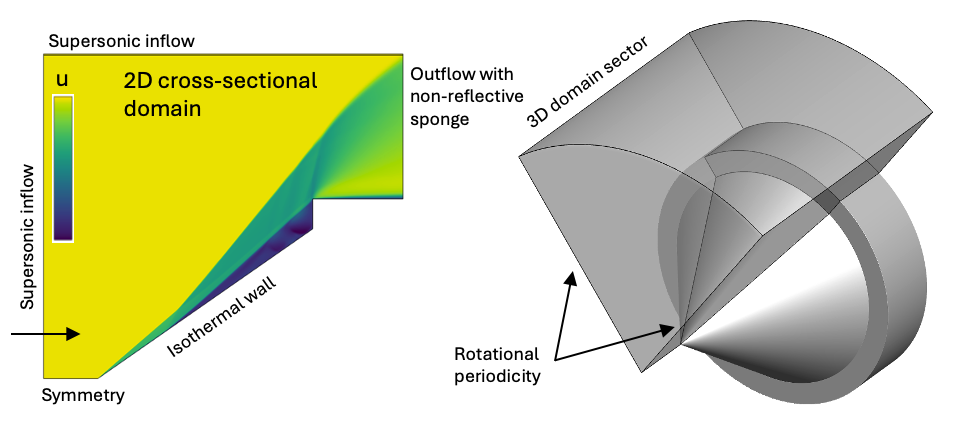}
    \caption{Computation domain and the boundary conditions for companion simulations. The 3D periodic domain extent in the azimuthal direction is one of the few different ranges considered in the study. The 3D sector domain resolves 3D small-scale fluctuations in the unsteady shear layer and the axisymmetric shock motions.}
    \label{fig:cfd_domain}
\end{figure}

%\rev{R3-4: axial singularity}
 The computational domain for the simulation is illustrated in Figure~\ref{fig:cfd_domain}. Both axisymmetric (2D) and three-dimensional (3D) computations are carried out to investigate the flow dynamics, assess the impact of non-axisymmetric effects, and validate the numerical approach against experimental observations. 
%Key aspects of the simulation setup are detailed below:

Axisymmetric (2D) simulations are first conducted for the steady cases. For the unsteady oscillatory cases,  simulations on the 3D sector domain are then carried out to capture azimuthal instabilities, secondary flows, and turbulent structures that cannot be captured in 2D.

\upd{ The 3D computational domain uses a 72\textdegree~azimuthal sector with rotational periodic boundary conditions. The sensitivity of the results to the azimuthal domain size are discussed in the Appendix.} A finite sector is employed in the azimuthal direction because the experiments exhibit axisymmetric shock dynamics. Therefore, only small-scale structures in the shear layer require azimuthal resolution. This enables higher computational resolution for capturing the small-scale 3D structures in the shear layer and near reattachment. 
%A sensitivity study showed that increasing the azimuthal extent to 64\textdegree~did not significantly affect separated flow statistics, and justifies the sector width within the scope of this study, providing sufficient azimuthal resolution.

%\adw{R2-5: provide grid resolution details} 
The finest mesh consists of approximately 25M cells for the 3D case and 200k cells for the 2D case, with clustering near critical flow regions such as boundary layers and shear layers. A grid independence study was performed by doubling the mesh resolution in x and $\theta$ direction. The shock system and shear layer angle was found to be unaffected. The near-wall spacing is $10^{-3}$ mm and is sufficient for obtaining $y^+ < 1$, enabling accurate resolution of the laminar boundary layer pre-separation.

Simulations are advanced at CFL of at most 5, until physical time of 0.5 ms to avoid transients. For oscillatory and pulsatory cases, the flow is initiated with flow-field extracted from lower \textcolor{blue}{Re$_L$} and is simulated up to 20 oscillatory cycles to collect data for frequency analysis. Time-statistics are then collected by sampling every 1 $\mu$s. 

\subsubsection{Numerical discretization}
% We utilize a finite-volume scheme to solve equation~\eqref{eq:nse}. A stable low-dissipation scheme based on the 'kinetic-energy consistent' method developed by ~\cite{subbareddy2009fully} is used for the  convective flux evaluation. This is a hybrid central/upwind scheme where the inviscid flux is split into a symmetric (non-dissipative) and an upwind (dissipative) portion. The upwind part is multiplied by a shock detector to ensure that dissipation is only active at shocks. A gradient reconstruction method is used to increase the accuracy of the central portion of the flux to (formally) sixth-order in smooth regions of the flow.
We utilize a finite-volume scheme to solve equation~\eqref{eq:nse}. The convective flux is evaluated using a stable low-dissipation scheme based on the kinetic-energy consistent (KEC) method developed by~\citet{subbareddy2009fully}. The KEC fluxes employ a hybrid central/upwind approach, where the inviscid flux is decomposed into symmetric (non-dissipative) and non-symmetric (dissipative) components of the modified Steger–Warming flux-vector splitting scheme~\citep{maccormack2014numerical}.  To enhance the accuracy of the central flux component in smooth flow regions, a gradient reconstruction technique is implemented, resulting in a formally sixth-order accurate scheme~\citep{Sidharth2018}. \textcolor{blue}{The upwind portion is modulated by a shock detector, ensuring that dissipation is localized to regions with strong compression and shocks.
The Ducros sensor \cite{ducros2000high} is used for shock detection and helps preserves small-scale turbulent structures in the separation zone.
} The viscous fluxes are computed using second-order least-squares gradients. Additional details can be found in \citet{candler2015development}. 

Time integration is carried out using a second order accurate implicit scheme~\citep{martin2006parallel},
\begin{align}
\label{eq:timeintIntro}
    \frac{3 \b{U}^{n+1}-4 \b{U}^n+\b{U}^{n-1}}{2 \Delta t} \,+\, \frac{\partial \b{F}_j^{n+1}}{\partial x_j} \,+\, \frac{\partial \b{F}_j^{v,n+1}}{\partial x_j} =0,
\end{align}
to be solved for solution at future time step $n+1$. The flux vector is linearized, 
\begin{align}
\label{eq:TimeIntLin}
\b{F}_j^{n+1} \simeq \b{F}_j^n 
\,+\, 
\left(\frac{\partial \b{F}_j}{\partial \b{U}_j}\right)^n\left(\b{U}_j^{n+1}-\b{U}_j^n\right) \;=\;
\b{F}_j^n
\,+\,
\b{A}^n_j \delta \b{U}_j^n
\end{align}
The Jacobian matrix $\b{A}^n_j$ is split, $
    \b A^n_{j}
    \;=\;
    \b A^n_{j+}
    \,+\,
    \b A^n_{j-}
$
corresponding to the positive and the negative wave characteristics of the flow. The solution for the next time step $\b{U}^{n+1}_{j}$ is solved iteratively as 
\begin{align}
\label{eq:time_update}
    \b U_j^{n+1}
    \;=\;
    \b U_j^p 
    \,+\,
    \b U_j^{p+1}
    \,-\,
    \b U_j^p
    \;=\;
    \b U_j^p
    \,+\,
    \delta \b U_j^p,
\end{align}
with an inner `$p$' iteration and an outer iteration for $\b U^{n+1}$. As $\delta \b U_j^p \to 0$, $\b U_j^p \to \b U_j^{n+1}$. The inner iteration loop is initialized using $\b U^{p} = \b U^{n}$. The iterative procedure to converge $\delta \b U_j^p \to 0$ can be found in \citet{martin2006parallel}.
% \begin{align}
% \label{eq:inner_outer}
% \begin{aligned}
% \frac{3}{2}\left(U_j^p+\delta U_j^p\right)-2 U_j^n+\frac{1}{2} U_j^{n-1} 
% = 
% & -\Delta t\left(\frac{\partial F_j}{\partial x_j} \,+\, \frac{\partial F^v_j}{\partial x_j}\right)^p \\
% &-\Delta t \frac{\partial}{\partial x_{j}}\left(A^p_{j} + A^p_{j,v} \right) \delta U_j^p
% \end{aligned}
% \end{align}
This time integration scheme has previously been utilized in 
carrying out high-fidelity simulations of transitional and turbulent hypersonic shock boundary layer interactions~\citep{dwivedi2022oblique}.

\subsection{Comparison with experimental data}

\subsubsection{\bf Steady cases}
\begin{figure}
    \centering
    \includegraphics[width=0.9\linewidth]{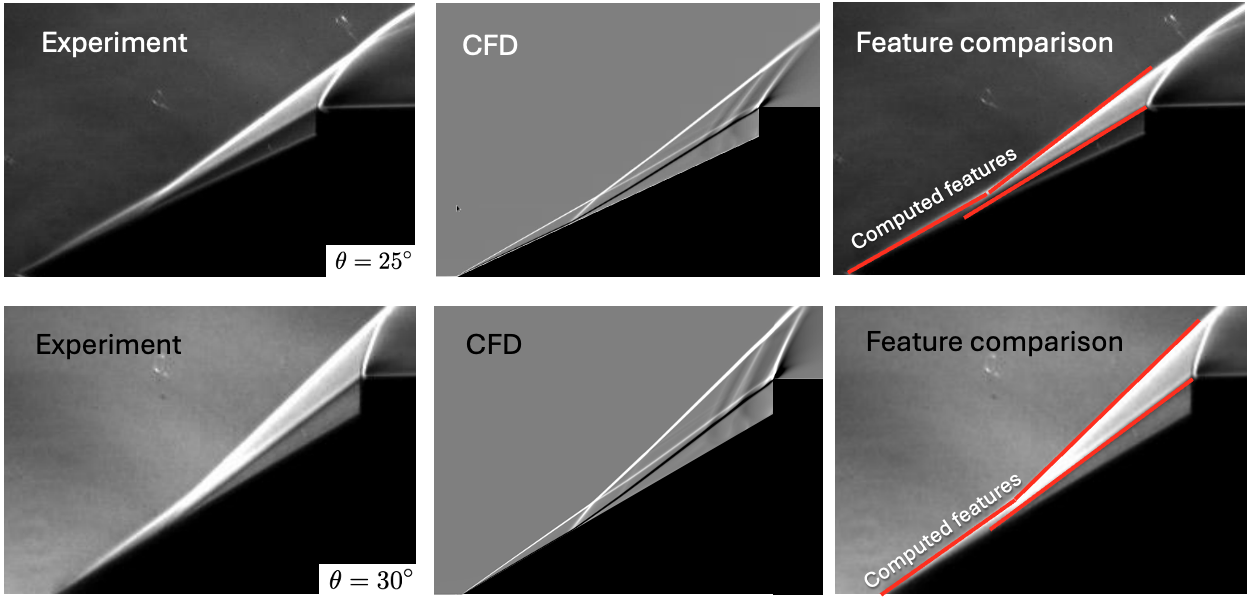}
    \caption{Feature (shock and shear layer) comparison between steady experiments ($\theta_1=25^\circ,30^\circ$) and axisymmetric 2D simulations. Cross-sectional $y-$density gradient from the simulation is used as the schlieren surrogate.}
    \label{fig:dg_comp_std}
\end{figure}

For steady cases, we conduct axisymmetric simulations for the three geometries considered in the experiment.
We compare the features of the simulated flow field against the experiments. This comparison is shown in Figure~\ref{fig:dg_comp_std}, where the nose-tip shock and separation shock extracted from the CFD simulations are plotted against the experimental data. As seen in the figure, these features align well with their experimental counterparts.
\upd{A quantitative comparison of the main flow features is presented in Table \ref{tab:steady_quant}.
This agreement serves as a validation of the separation location, the angle of the shocks and the shear layer before unsteadiness sets in.}

\begin{table}
{\color{blue}
  \centering
  \setlength{\tabcolsep}{8pt}
  \renewcommand{\arraystretch}{1.25}
  \begin{tabular}{@{}ccccccc@{}}
    \toprule
    \textbf{$\theta_1$ $(^\circ)$} & {Re$_L$} & \textbf{$x_s/L$} &
    \textbf{$\theta_c$ $(^\circ)$} & {$\theta_\text{sc}$ $(^\circ)$} & {$\theta_\text{SL}$ $(^\circ)$} \\
    \midrule
    & & Exp / CFD & Exp / CFD / TM & Exp / CFD & Exp / CFD \\
    \midrule
    25  & $3.9\times10^{5}$  & 0.36 / 0.36 & 30 / 30 / 29.4 & 37 / 37 & 31 / 31 \\
    25  & $7.5\times10^{5}$  & 0.31 / 0.31 & 29 / 29 / 29.4 & 37 / 36 & 31 / 31 \\
    30 & $3.3\times10^{5}$  & 0.36 / 0.36 & 35 / 35 / 34.9 & 44 / 44 & 37 / 37 \\
    35  & $2.9\times10^{5}$  & 0.27 / 0.27 & 40 / 40 / 40.6 & 48 / 50 & 42 / 42 \\
    \bottomrule
  \end{tabular}
  \caption{Comparison of experiment (Exp), CFD, and Taylor-Maccoll (TM) results for cone shock angle; experiment and CFD for separation shock and shear layer angles at selected cone and Re$_L$ values. All angles are reported in degrees.}}
  \label{tab:steady_quant}
\end{table}

\subsubsection{\bf Unsteady case: Shear layer fluctuations}

\begin{figure}
    \centering
    \begin{overpic}[width=1\linewidth]{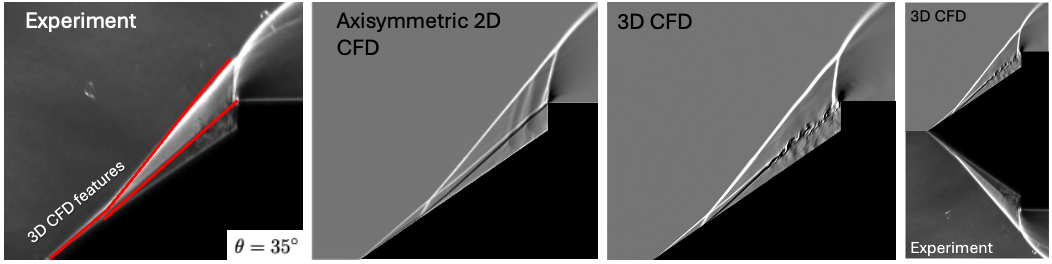}
     \put(23,10){\color{white}(a)}   
     \put(52,10){\color{white}(b)}   
     \put(80,10){\color{white}(c)}   
     \put(95,10){\color{white}(d)}   
    \end{overpic}
    \caption{Feature comparison between unsteady shear experiment and simulation $\theta_1=35^\circ$, \textcolor{blue}{Re$_L$}=2.9$\times10^5$. $y-$density gradient from the simulation is used as the schlieren surrogate. An axisymmetric 2D simulation is also shown for comparison, and underpredicts the separation zone.}
    %\sidgs{R-14: how is the experiment unsteady show..}}
    \label{fig:dg_comp_unstd}
\end{figure}

First, we conduct axisymmetric simulations for all conditions considered in the experiments.
As is observed in the experiments, an increase in the cone angle leads to an increase in flow unsteadiness. 
 The axisymmetric simulations in Figure~\ref{fig:dg_comp_unstd} (b) capture the onset of destabilizing waves within the separation bubble. However, shedding due to shear layer breakdown is not observed, even though Figure~\ref{fig:dg_comp_unstd} (a)  shows that the experimental results exhibit unsteady fluctuations in the shear layer.
 Therefore, we carry out three-dimensional simulations for all the conditions in the experiments. We compare the 3D simulation results with the experiments in Figure~\ref{fig:dg_comp_unstd} (c-d). In these cases, the shear layer breaks down in a manner similar to the experiments, resulting in unsteady fluctuations post-interaction with the contact wave. A detailed stability analysis of this phenomena is presented in section~\ref{sec:lst_contact}.  An important difference between the axisymmetric and 3D simulations is the separation location. We find that in the 3D simulations, the separation point is further upstream compared to the axisymmetric case and matches the experimental observation. For the $35^\circ$ case at \upd{ {Re$_L$} = 2.9$\times10^5$, the separation location corresponds to approximately $0.3L$.}

We also analyze the nature of shear layer reattachment in the 3D versus axisymmetric (2D) cases. In the 2D case, the shear layer reattaches directly at the step, whereas in the 3D case, pockets of entrained fluid prevent a well-defined mean reattachment line. 
%Additionally, a comparison of the mean 3D and 2D axisymmetric solutions reveals key differences, which we discuss in the following sections.

%Bench

% \begin{figure}
%     \centering
%     \includegraphics[width=1\linewidth]{30.png}
%     \caption{\adw{blue}{combine with the previous image}}
%     \label{fig:enter-label}
% \end{figure}

%\subsubsection{Flow unsteadiness}
\begin{figure}
    \centering
    \begin{overpic}[width=1.0\linewidth]{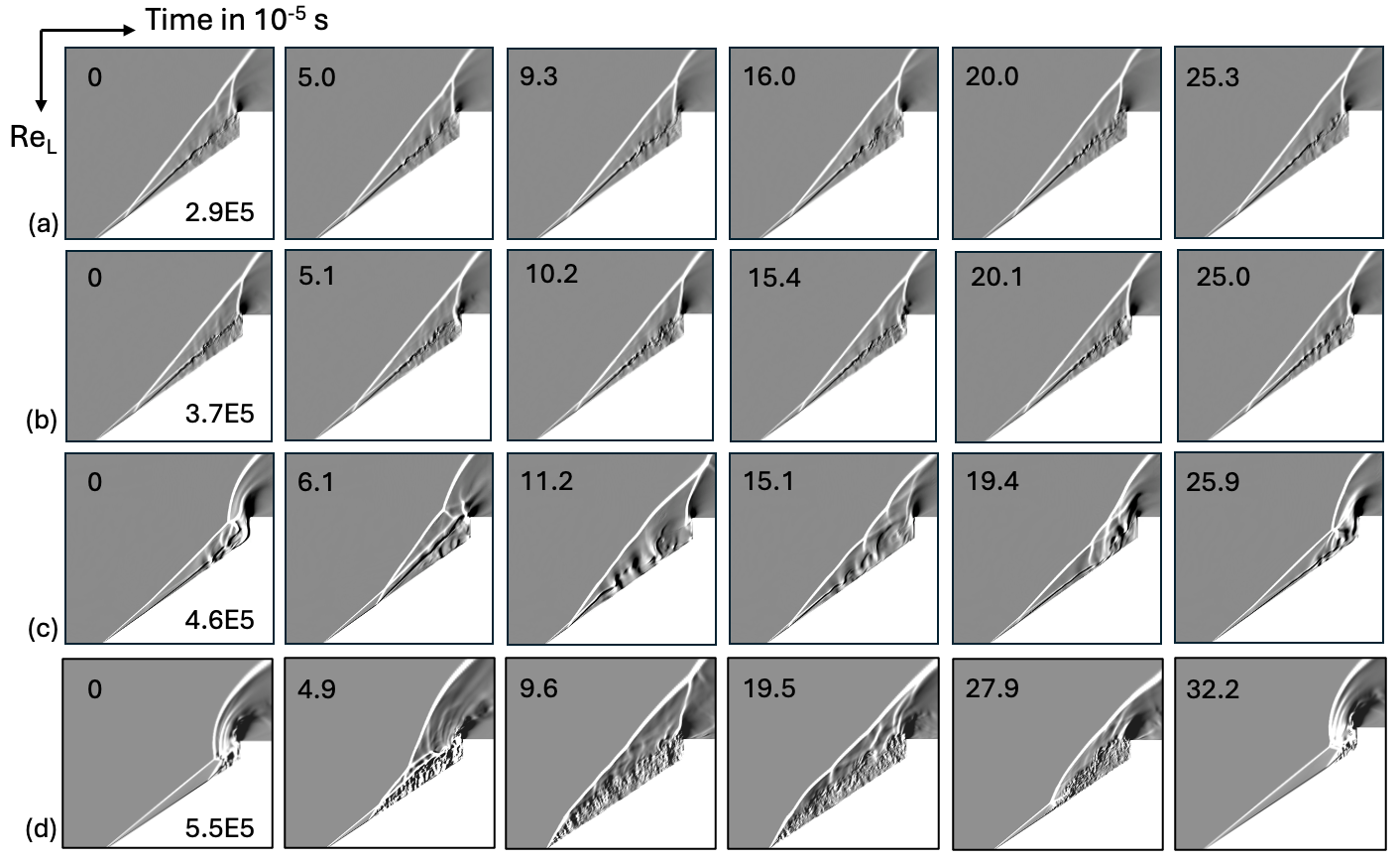}
     %    \put(5,4){(c)}   
     %    \put(5,19.7){(b)}   
     % \put(5,35){(a)}        
   \end{overpic}
    \caption{Simulation time frames for (a,b) shear layer, (c) oscillatory and (d) high-amplitude oscillatory (pulsatory) cycle unsteadiness observed with increase in \textcolor{blue}{Re$_L$} for the cone angle case of $35^\circ$.
    %\rev{R3-5: provide the Re7.7 pics}
    }
    \label{fig:Re_vs_unsteady}
\end{figure}

\begin{figure}
  \begin{overpic}[width=1\linewidth]{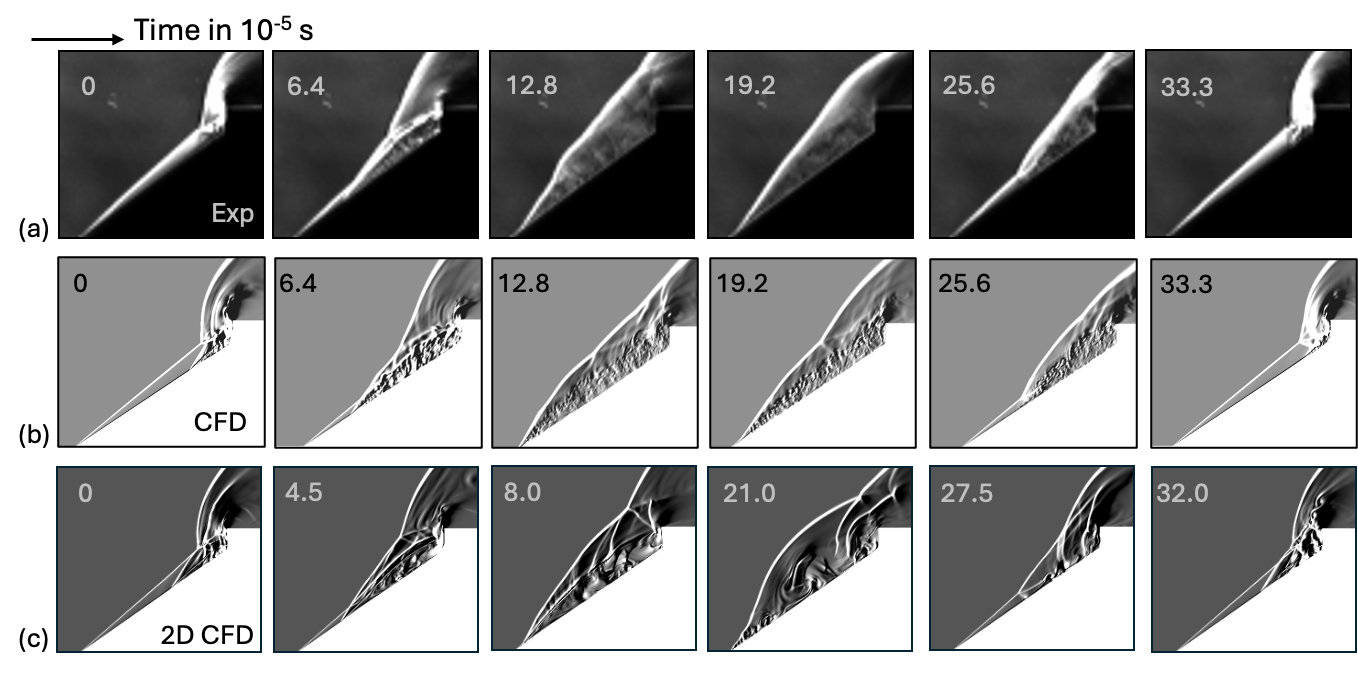}
    % \put(0,46.7){\vector(1,0){10}} % simple arrow
    % \put(11,4){ \small  {Time in $10^{-5}$s}} % label
    \end{overpic}
  \caption{Frames of pulsatory cycle compared against experimental images for the $\theta_1=35^\circ$, \textcolor{blue}{Re$_L$}=5.5$\times10^5$ case. Results from axisymmetric CFD are also shown. The pulsatory shock dynamics is approximately axisymmetric but is affected by the 3D shear layer. 
  Please see the supplementary movie file for experimental schlieren. 
  %\rev{R3-7}
  }
\label{fig:puls_axisym}
\end{figure}

%Separation unsteadiness 
% \begin{figure}
%     \centering
%     \includegraphics[width=1\linewidth]{unsteady_sbli_comparison.png}
%     \caption{Frames of pulsatory cycle compared against experimental images for the $\theta_1=35^\circ$, Re=15M case. Results from axisymmetric CFD are also shown. The pulsatory shock dynamics is approximately axisymmetric but is affected by the 3D shear layer. }
%     \label{fig:puls_axisym}
% \end{figure}

\subsubsection{\bf Unsteady case: Separation zone oscillations}

We compute the three-dimensional flow fields for all \textcolor{blue}{Re$_L$} cases at $\theta_1 = 35^\circ$. Figure~\ref{fig:Re_vs_unsteady} shows snapshots of the density gradient in a plane, at different time instances during the cycle, for the different \textcolor{blue}{Reynolds number} cases. At \textcolor{blue}{Re$_L$} = 3.7$\times10^5$, the shear layer is three-dimensional and associated with unsteady fluctuations, while the separation zone remains anchored. However, with an increase in \textcolor{blue}{Re$_L$} to 4.6$\times10^5$, the separation shock and the recirculation bubble begin to oscillate along the length of the cone. The reattachment shock in front of the step also moves upstream intermittently, followed by bursts of bubble shedding.

% \subsubsection{\bf Unsteady case: Separation zone pulsations}
With further increase in \textcolor{blue}{Re$_L$}, at \textcolor{blue}{Re$_L$} = 5.5$\times10^5$, the strength of the oscillations increases, with the separation shock traversing all the way upstream toward the nose tip during the oscillation cycle. These oscillations are referred to as ``pulsations''. The time-history of the flowfield during a pulsation cycle is further compared to experimental schlieren snapshots in Figure~\ref{fig:puls_axisym}. 
%To investigate the role of three-dimensionality in pulsation cycle we compare the flow at different instances during the pulsation cycle with both two-dimensional axisymmetric and three-dimensional simulations. 
For comparison, an axisymmetric simulation is also compared at this \textcolor{blue}{Reynolds Number}. As shown in the figure, the pulsating unsteadiness is also observed in the axisymmetric simulations, although the separated flow structures are not physical. This observation suggests that the shock oscillation phenomenon is inherently axisymmetric, modulated by the 3D fluctuations in the separation-reattachment zone.

Next, we discuss the unsteady flowfield during the oscillation and pulsation cycles in detail.

\subsection{Unsteady shock dynamics during an oscillatory/pulsatory cycle} 
\subsubsection{\bf Oscillation cycle}
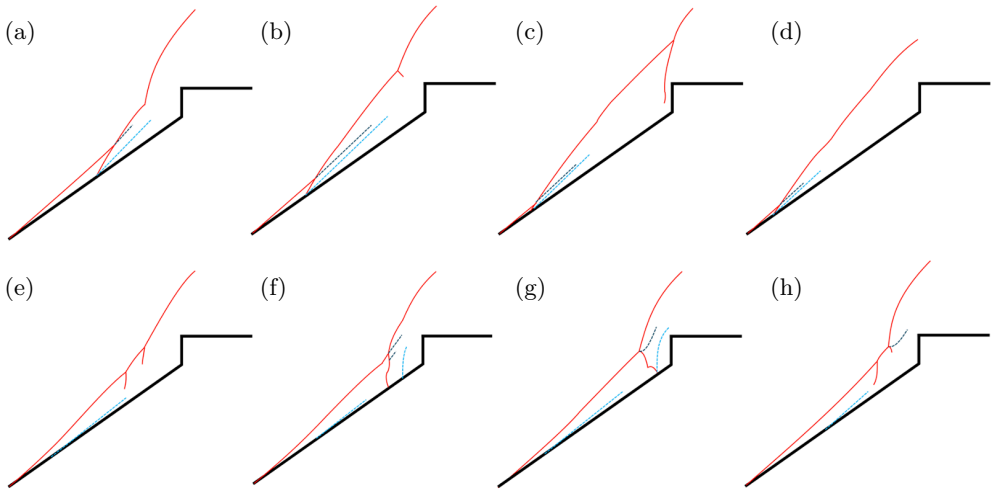
\begin{figure}
  \centering
  \begin{overpic}[width=\textwidth]{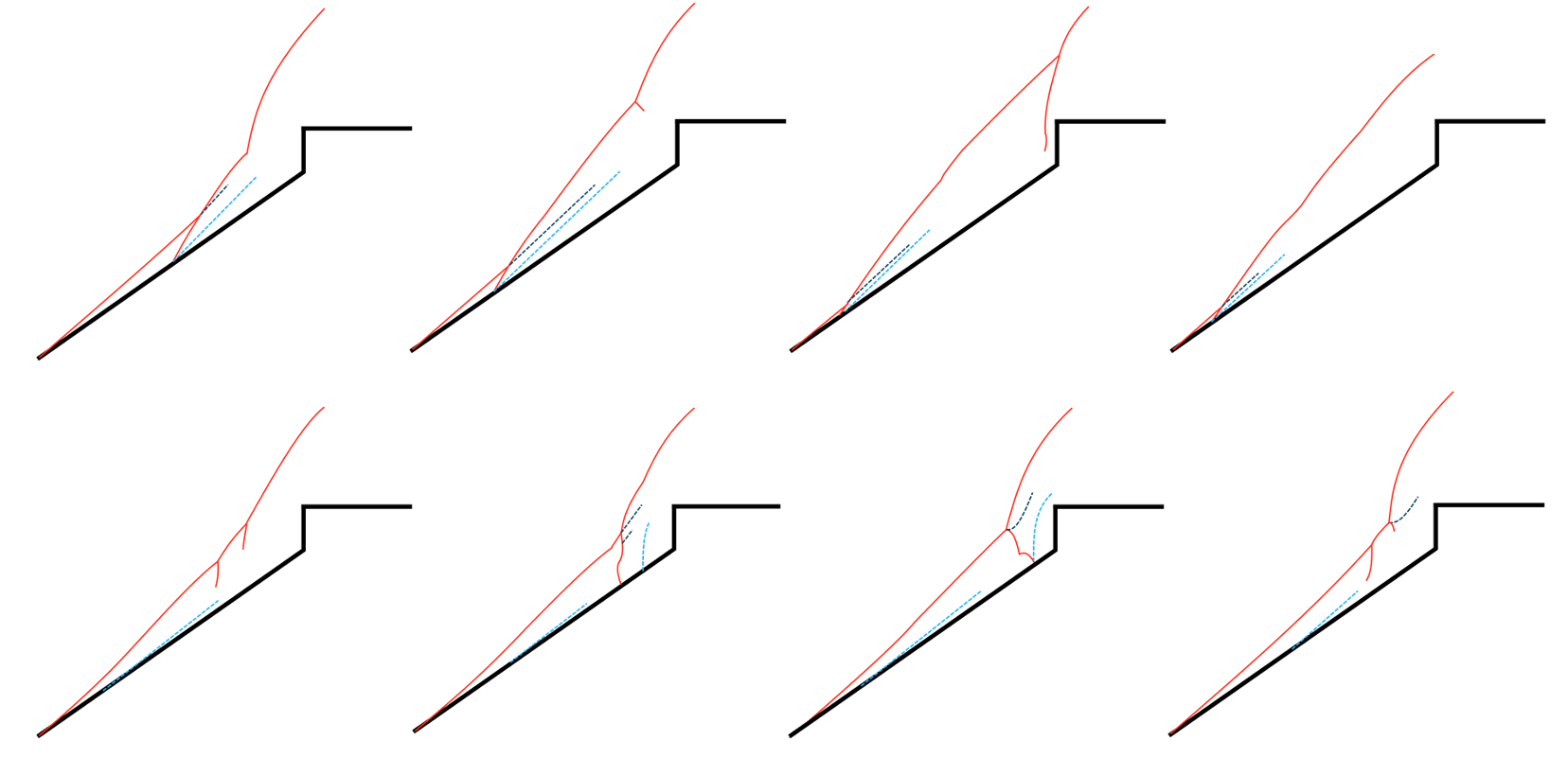}
 % Top row, from left to right - will replace existing (e)(f)(g)(h)
    \put(2,45){(a)}
    \put(27,45){(b)}
    \put(52,45){(c)}
    \put(77,45){(d)}
    
    % Bottom row, from left to right
    \put(2,20){(e)}
    \put(27,20){(f)}
    \put(52,20){(g)}
    \put(77,20){(h)}
  \end{overpic}
  \caption{%\adw{R2-8: describe the qualitative differences between 12 and 15} 
  Shock-shock interactions during the oscillatory unsteadiness cycle for \textcolor{blue}{Re$_L$}=4.6$\times10^5$. The red line indicates shock, the  dark blue dashed lines indicates the contact from the shock-shock interaction and the light blue dashed line indicates the shear layer associated with the recirculation.}
  \label{fig:ss_interac_osc}
\end{figure}

As discussed, in the high-\textcolor{blue}{Re$_L$} regime, the flow exhibits strong shock-boundary layer interactions and pronounced unsteady separation dynamics that give rise to rapidly evolving shock wave patterns. We extract the shocks, the contact waves and the incipient separated boundary layer in the large amplitude oscillatory regime before transitioning into the pulsatory unsteadiness.  This is illustrated in Figure~\ref{fig:ss_interac_osc} for \textcolor{blue}{Re$_L$}=4.6$\times10^5$ for one cycle of the oscillation. 

We choose a point in the cycle when the separation point starts to shift upstream, and the associated recirculation zone expands.
Upon reaching its maximum upstream extent, the separation shock disappears, and the recirculation bubble collapses abruptly.
The mass of the fluid from the collapsed bubble is shed downstream forming a normal shock near the step.
As this fluid accumulates near the step, the normal shock propagates outward and transforms into an outward moving oblique separation shock, accompanied by a corresponding upstream motion of the separation point.
This sequence repeats cyclically, defining the unsteady shock-separation interaction that governs a large amplitude oscillatory regime.

At \textcolor{blue}{Re$_L$}=5.5$\times10^5$, the oscillations of the separation shock and the bow shock become larger and a `pulsatory' unsteady shock–boundary layer interaction emerges.

\subsubsection{ \bf Pulsation cycle}
\begin{figure}
    \centering
    \begin{overpic}[width=1\linewidth]{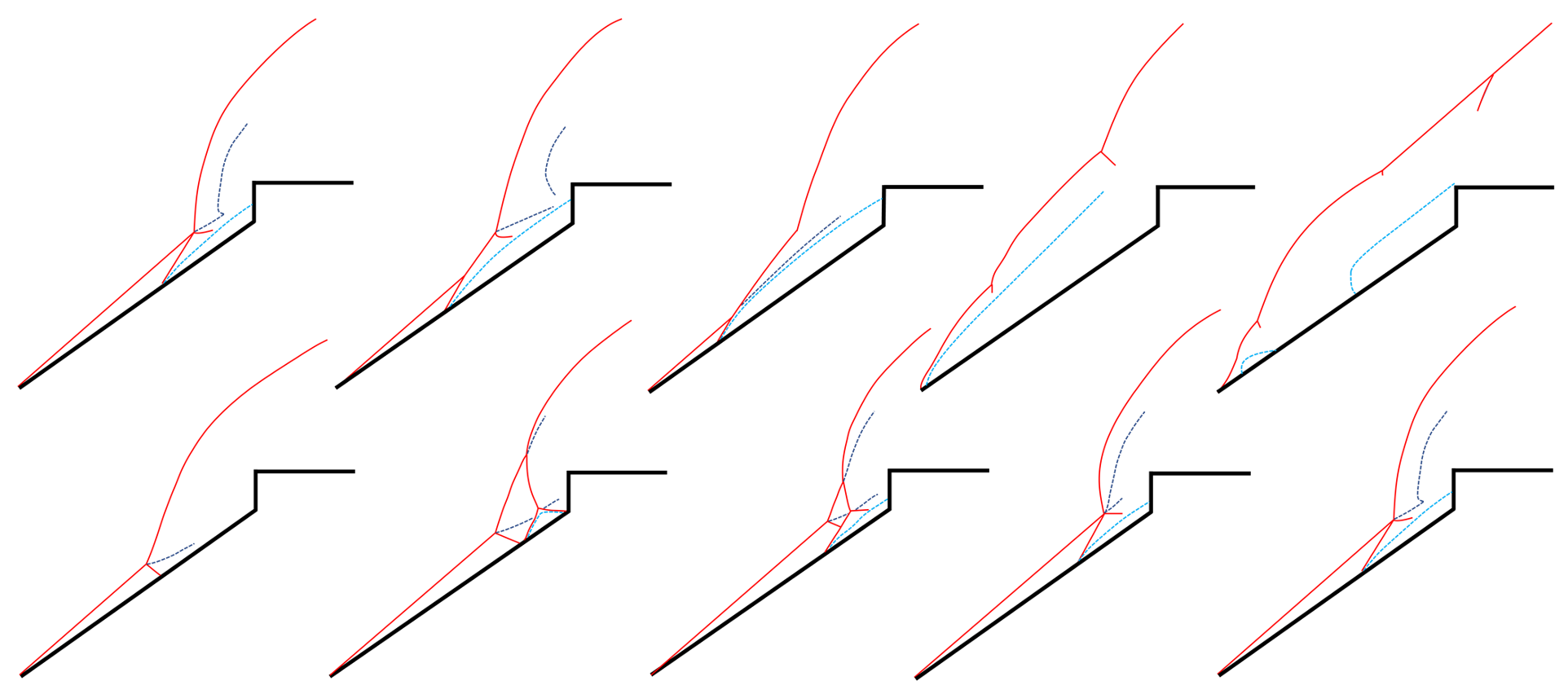}
    \put(3,40){(a)}
    \put(25,40){(b)}
    \put(47,40){(c)}
    \put(67,40){(d)}
    \put(86,40){(e)}

    % Bottom row (5 columns)
    \put(3,15){(f)}
    \put(25,15){(g)}
    \put(47,15){(h)}
    \put(67,15){(i)}
    \put(86,15){(j)}
    \end{overpic}
    \caption{Shock-shock interactions during the oscillatory unsteadiness cycle for \textcolor{blue}{Re$_L$}=5.5$\times10^5$. The red line indicates shock, the  dark blue dashed lines indicates the contact from the shock-shock interaction and the light blue dashed line indicates the shear layer associated with the recirculation.}
    \label{fig:ss_interac_puls}
\end{figure}
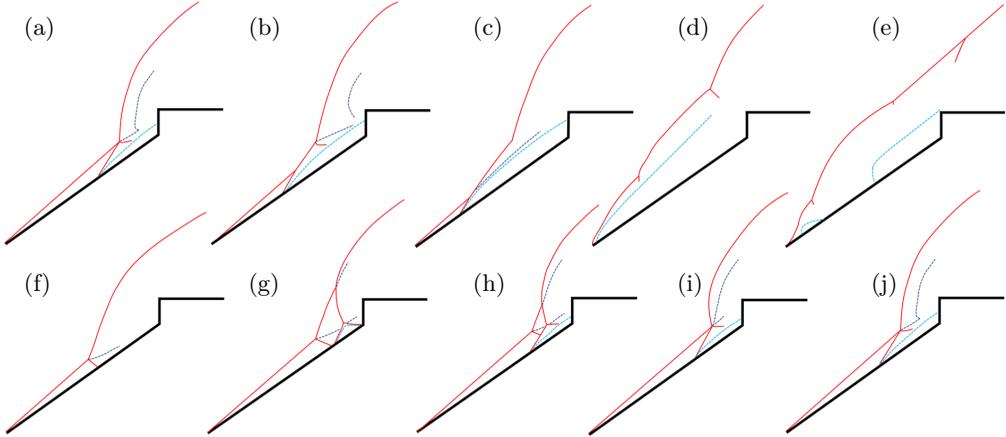

At higher \textcolor{blue}{Reynolds numbers}, the \textit{pulsatory} unsteadiness sets in, as illustrated in Figure~\ref{fig:ss_interac_puls} for \textcolor{blue}{Re$_L$}=5.5$\times10^5$. This pulsation cycle has previously been described in detail in the work of \cite{Ozawa2009,Wang2011}, in which an $M=3,\textcolor{blue}{Re_L=4.7\times 10^6}, \theta_1=20^\circ, L/D=0.985$ cone-step flow is considered.

Figure~\ref{fig:ss_interac_puls} provides an overview of the dynamics during the pulsation cycle.
The first frame in Figure~\ref{fig:ss_interac_puls} corresponds to an instant when the separation zone begins to grow and propagate upstream. A near-vertical contact, a remnant of a transient lambda shock - bow shock interaction is visible. 
This feature is also evident in the experimental schlieren image (see Figure~\ref{fig:exp_para_schl}).

As the separation shock moves upstream and the separation zone continues to expand, the bow shock weakens. Eventually, the leading-edge shock and the separation shock coalesce and the flow separates very close to the nose tip. Finally, the separation bubble collapses and sheds downstream. % similar to a starting vortex. 

A new boundary layer begins to develop along the cone surface after separation collapse, and it advances downstream with a Mach stem - lambda shock system \citep{Combs2018} instead of an oblique shock. The Mach stem in the pulsation cycle is also observed in the experimental images (see Figure ~\ref{fig:exp_para_schl}), and has previously also been seen in the work of \cite{Ozawa2009,Wang2011,Wang2012}. As fluid mass accumulates near the step, a new separation zone emerges, accompanied by the formation of a new oblique separation shock, which propagates upstream and lifts the Mach stem off the surface. The sequence repeats, defining the characteristic behavior of the pulsatory regime at high \textcolor{blue}{Reynolds numbers}. For the present conditions, each cycle lasts about 0.3 ms, matching the flow-through time.

\upd{
As pointed out by a reviewer, we must 
note that while the unsteady shock structures appearing over the course of the oscillation cycles are different at the two Reynolds numbers, the underlying unsteadiness mechanism is qualitatively similar. The higher Reynolds number simply leads to larger amplitudes and correspondingly different instantaneous shock shapes are manifestations of the response to a similar driving mechanism.
}

% \begin{figure}
%     \centering
%     \includegraphics[width=0.85\linewidth]{shock_sketch_Re12_2.png}
%     \caption{Shock-shock interactions during the oscillatory unsteadiness cycle for Re = 12.5M. The red line indicates shock, the  dark blue dashed lines indicates the contact from the shock-shock interaction and the light blue dashed line indicates the shear layer associated with the recirculation.}
%     \label{fig:ss_interac_osc}
% \end{figure}

\subsection{SPOD of numerical simulation data}

Similar to the spectral proper orthogonal decomposition (SPOD) analysis performed on the experimental schlieren data presented in Section~\ref{sec:exp_spod}, we conducted a corresponding SPOD analysis on the numerical simulation results (details in Appendix \ref{spod_details}) The objective was to characterize and compare the dominant modes from the computed fields to those obtained from the experimental schlieren intensity images.

%identify the distribution of energy across frequencies and modes, and determine the dominant frequency associated with the inherent flow unsteadiness.

\begin{figure}
    \centering
   \includegraphics[width=1\linewidth]{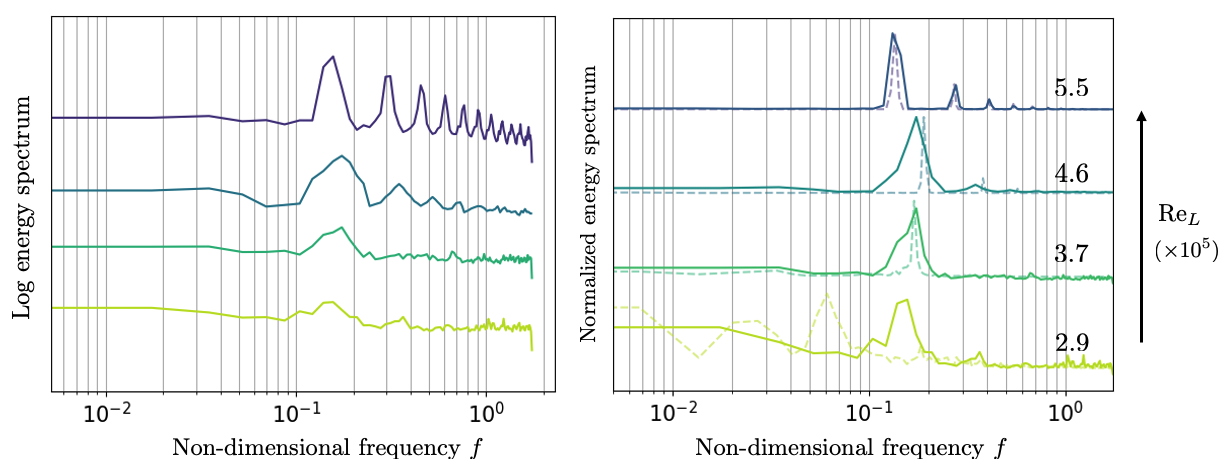}

    \caption{SPOD spectrum from 3D simulation data (solid lines) for $\theta_1=35^\circ$. Comparison to SPOD spectra from experimental schlieren image is shown with a dashed line. Please note that the spectra for each case is arbitrarily shifted on the y-axis.}

    \label{fig:spod_cfd_exp_comp}

\end{figure}

This analysis was performed on simulation data obtained for all \textcolor{blue}{Re$_L$} cases at $\theta_1=35^\circ$. The SPOD energy spectra (energy content as a function of frequency) are presented in Figure~\ref{fig:spod_cfd_exp_comp}. As observed, the dominant frequency associated with the flow unsteadiness is approximately constant across \textcolor{blue}{Re$_L$}. 

The Strouhal number of the dominant SPOD mode from the simulation data is calculated to be St $\approx$ 0.17. 
Except for the highest \textcolor{blue}{Re$_L$} case, this is in close agreement with the value derived independently from the SPOD analysis applied to the experimental schlieren intensity fields (detailed in Section~\ref{sec:exp_spod}). Interestingly, the analysis shows that the low frequency unsteadiness associated with the dominant mode can be observed in simulations even in the lowest \textcolor{blue}{Reynolds number} case \textcolor{blue}{Re$_L$} = 2.9$\times10^5$. This is in contrast to the experiments where the shock motion is small and the schlieren intensity field does not have enough signal-to-noise ratio to identify this mode. 

Furthermore, a key contribution of the numerical study is the confirmation of the onset and nonlinear amplification of flow separation-induced unsteadiness as a function of increasing \textcolor{blue}{Reynolds number}. For the highest \textcolor{blue}{Re$_L$} case, the nonlinear harmonics associated with the primary mode are also distinctly observed, consistent with the experiments. This consistency between the numerical and experimental findings provides validation for the simulation methodology and the quiet freestream conditions of the tunnel alike.

\subsubsection{\bf Spatial imprint of the SPOD mode}

To facilitate a direct comparison between the numerical results and the experimental flow visualizations, synthetic plane schlieren images were computed from the simulation data. These images were generated based on density gradients ($\nabla \rho$) calculated from the SPOD modes on the center-plane within the computational domain.

%\adw{R2-9: report new results} 
The spatial structures associated with the dominant mode identified from the simulations are compared against the corresponding leading modes obtained from the SPOD analysis of the experimental schlieren data. It is important to recognize that experimental schlieren techniques typically provide intensity-based measurements, which are inherently line-of-sight integrated quantities and thus cannot directly compare to the planar information available from the numerical simulations. Therefore, both techniques provide complementary analyses of the flowfield.

\begin{figure}

    % \centering
    \centering 
    \includegraphics[width=1\linewidth]{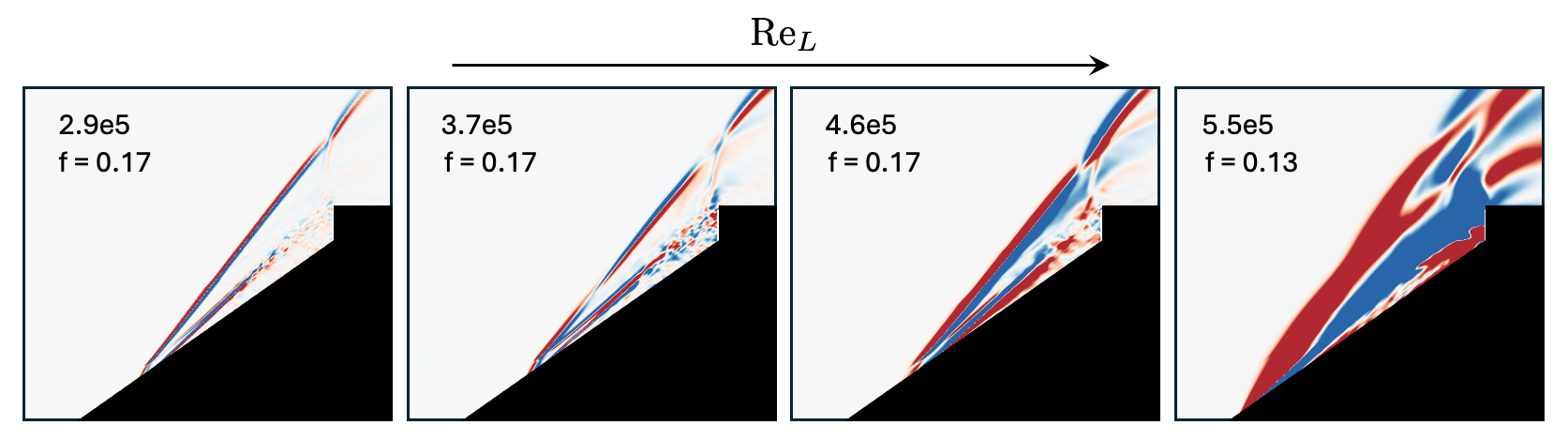}
    \caption{Vertical density gradients at the centerline azimuthal plane of the leading SPOD mode from simulations are shown.  $f = f^*L/U_\infty$ denotes the non-dimensional frequency of the modes.}
    %(CFD) and experimental schlieren intensity (Exp) for $\theta_1=35^\circ$. Two \textcolor{blue}{Re$_L$} cases, \textcolor{blue}{Re$_L$} 3.7$\times10^5$ (low-amplitude oscillatory) and 5.5$\times10^5$ (high-amplitude pulsatory) cases are shown. $f = f^*L/U_\infty$ denotes the non-dimensional frequency. \rev{R3-6: mode shape at Re 7.7}, \rev{R3-8}}

    % \caption{Comparison between the spatial structure of the leading SPOD mode from simulation vertical density gradient (CFD) and experimental schlieren intensity (Exp) for $\theta_1=35^\circ$. Two \textcolor{blue}{Re$_L$} cases, \textcolor{blue}{Re$_L$} 3.7$\times10^5$ (low-amplitude oscillatory) and 5.5$\times10^5$ (high-amplitude pulsatory) cases are shown. $f = f^*L/U_\infty$ denotes the non-dimensional frequency. \rev{R3-6: mode shape at Re 7.7}, \rev{R3-8}}

    \label{fig:spod_mode_comparison}

\end{figure}

A comparison of the leading SPOD modes, as illustrated in Figure~\ref{fig:spod_mode_comparison}, demonstrates a reasonable qualitative agreement. Key dynamic features associated with the flow unsteadiness are captured similarly in both datasets. Specifically, the characteristic motion of the primary separation shock and the downstream reattachment bow shock system, including their relative phasing, is clearly represented within the dominant mode structures from both simulation and experiment.
% Observations of the mode structures suggest that the shear layer region appears somewhat noisy or less defined in the SPOD modes. This is attributed to aliasing effects stemming from the temporal sampling rate used for the simulation output for the multiscale fluctuations. However, this shear layer activity does not represent the dominant contribution to the energy content of the leading modes identified. 
\upd{While structures in shear breakdown region are also observed, they are weak compared to the modal imprint corresponding to the shock motion.}

%An important aspect of the analysis is the finding that similar to the experiments, no resonant traveling acoustic waves are identified in the dominant, most energetic SPOD modes derived from the simulation data within the frequency range studied. This suggests that the primary unsteadiness mechanism captured by SPOD at St $\approx$ 0.17 is predominantly hydrodynamic in nature, rather than being driven by any resonant acoustics propagating between the separation shock and the step (see Appendix).

% \subsection{Analysis and discussion}

% We outline the novel results from the present study:

% \begin{itemize}

% \item Modal dynamics in the unsteady shear layer case

% \item Insight into the oscillatory shock-boundary layer interaction: 

% \end{itemize}

% - Understand the different phases in the cycle

% - Comment on the state of the flow

% - Shock-interaction sketches

% - Focus on forward and backward movement of separation

% - Mass flux and surface properties

\subsection{Three-dimensional features in the separation zone during unsteadiness: Insight from the dominant SPOD Mode}

In addition to the dominant axisymmetric dynamics, the spectral proper orthogonal decomposition (SPOD) modes are further analyzed to identify three-dimensional flow structures within the unsteady flow, particularly during the low amplitude oscillation regime i.e. mid-\textcolor{blue}{Re$_L$} cases. This analysis focused specifically on the leading (most energetic) SPOD mode at \textcolor{blue}{Re$_L$}=4.6$\times10^5$ which corresponds to the onset conditions for significant flow unsteadiness.

A key observation for this leading mode is the presence of a distinct non-zero velocity component in the azimuthal direction ($u_\theta$). The component being non-zero definitively indicates that the primary instability mechanism, captured by the dominant SPOD mode, possesses a significant three-dimensional character. While it is well-understood that shear layers inherently develop three-dimensional structures as part of their natural transition process~\citep{Brown1974, Bradshaw1977, Papamoschou1988}, the key finding here is that the dominant mode's three-dimensionality is prominently located within the separation zone itself, and not just confined to the outer shear layer.

\begin{figure}

    \centering

    \includegraphics[width=0.65\linewidth]{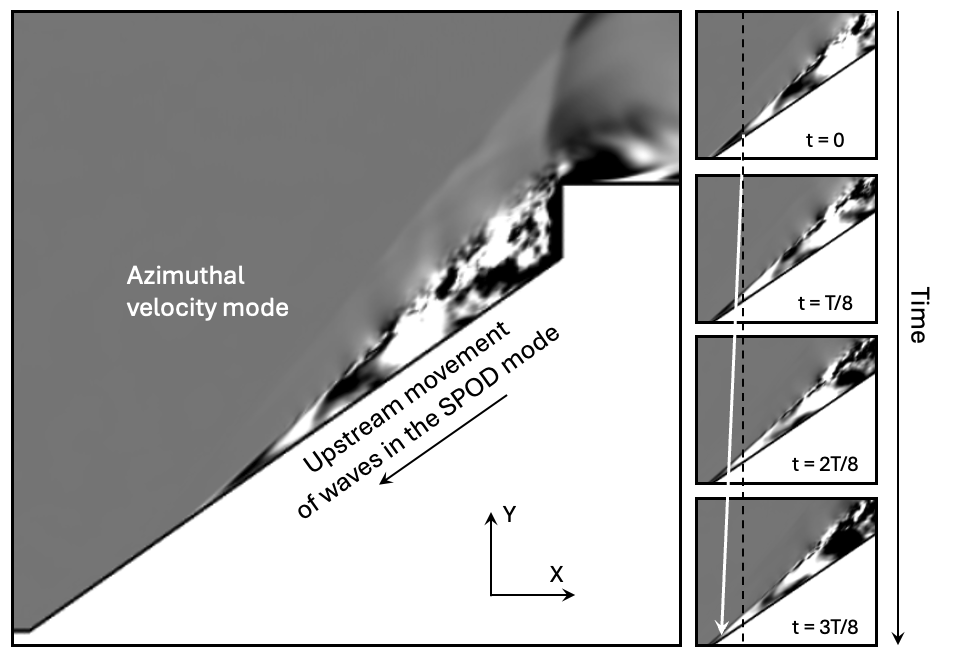}

    \caption{Azimuthal velocity component of the leading SPOD mode from simulations for the low-amplitude oscillatory case $\theta_1 = 35^\circ$ and \textcolor{blue}{Re$_L$} = 3.7$\times10^5$. Insets show the direction of the group velocity associated with the three dimensional mode at different instants in a time-period.}

    \label{fig:up3d}

\end{figure}

The spatial organization of these three-dimensional features is illustrated through visualizations of the instantaneous fluctuations from the unsteady SPOD mode. Figure~\ref{fig:up3d} presents the $u_\theta$ component of the leading mode with time-snapshots of the instantaneous three-dimensional flow field. As depicted in these figures, the $u_\theta$ waves travel upstream, with the local reversed flow within the separation zone. The modal analysis points towards predominantly hydrodynamic perturbation waves residing within the separation bubble (as opposed to traveling acoustic waves outside the bubble, as is discussed in the Appendix). This finding is consistent with analyses reported in the literature concerning intrinsic instabilities in high-speed separated flows \cite{gs2018onset,Cao2021,Cao2022}. A quantitative estimate of the strength of the three-dimensionality is given by the characteristic magnitude of the azimuthal velocity fluctuations associated with the dominant mode.  It is found to be approximately $|u_\theta|_{max} \approx 0.2 U_\infty$, where $U_\infty$ represents the freestream velocity, indicating a non-negligible value being present even at the early stages of the flow unsteadiness.

\section{Spectral submanifold analysis of unsteadiness}

% To better understand the onset of unsteadiness, we analyze the Kelvin–Helmholtz (KH) instability in the presence of a separation shock–nose tip shock interaction. The contact surface emanating from this interaction intercepts the separated shear layer and induces unsteady fluctuations, which appear critical to the onset of unsteadiness, as observed in both experiments and simulations. The analysis in this section provides quantitative support for this hypothesis. 
For oscillatory and pulsatory unsteadiness, we perform Spectral Submanifold (SSM) analysis~\citep{Cenedese2022}, as the dynamics exhibit a limit cycle. This allows us to infer the nature of the underlying linear and nonlinear components driving the observed behavior. The nonlinear model technique helps identify the possibility of existence of a supercritical Hopf bifurcation in the flow system.

% - To better understand the onset of unsteadiness, we analyse the KH instability in the presence of separation shock -nose tip  shock interaction. The contact emanating from the interaction intercepts the separated shear layer and leads to unsteady fluctuations.  This appears critical to the onset of unsteadiness from both the epxierments and the simulations. The analysis in this sections provides quantitiative credence to this hypothesis.
% - For oscillatory and pulsatory unsteadiness, we carried out SSM because dynamics has a limit cycle already. And back infer the natyre of the linear component and the nonlinear component. 

% ADD INTRODUCTION TO THIS SECTION HERE!

%\subsection{Role of contact discontinuity: linear stability of variable density shear layers}
%\adw{blue}{add LST results here }

% \subsection{Nonlinear model reduction: presence of supercritical Hopf bifurcation}
\label{sec:SSM}

% As the \textcolor{blue}{Re$_L$} increases, separation unsteadiness leads to stronger oscillations. To capture such growing nonlinear affects, 
We analyze the time-series data of the integral flow quantities  such as the mass-flux. We develop a nonlinear reduced-order model (ROM) that describes transition from small to large amplitude oscillations in mass flux for a given geometry as Reynolds number increases. %As the Re increases, separation unsteadiness leads to stronger oscillations in integral flow quantities. To capture the growing nonlinear effects, we analyze time-series data of the mass flux and develop a nonlinear reduced-order model (ROM) that describes the transition from small to large amplitude oscillations for a fixed geometry. 
We use Spectral Submanifold (SSM) analysis~\citep{Jain2022,Cenedese2022} to extract ROMs near attractors, which is well-suited for the observed limit-cycle behavior. This allows us to separate and infer both linear and nonlinear contributions to the unsteady flow dynamics.

% \subsubsection{\bf Spectral submanifold (SSM) approximation}
%The onset of separation unsteadiness introduces time dependence in bulk flow quantities. Once the critical Reynolds number Re is exceeded, simulations and experiments show that the amplitude of the oscillations increases with Re. To quantify the separation unsteadiness dynamics for the cone-step geometry under investigation, we analyze time-series data of non-dimensionalized mass flux integrated across a vertical plane $m(t)$ at  $x/L = 0.5$, where $L$ is cone length. This parameter is sensitive to the breathing and collapse of the separation zone. A similar choice was made to study the influence of free-stream enthalpy on shock-boundary layer interactions over a double-wedge configuration \cite{reinert2020simulations}. 

The onset of separation-induced unsteadiness introduces temporal oscillation in bulk flow quantities. Beyond a critical \textcolor{blue}{Re$_L$}, both experimental and numerical studies reveal a monotonic increase in the amplitude of these oscillations with increasing \textcolor{blue}{Re$_L$}. To characterize the unsteady separation dynamics in the cone-step geometry, we examine time-resolved measurements of the non-dimensionalized mass flux, $\dot{m}(t)$, spatially integrated across a vertical plane located at $x/L = 0.5$, where $L$ denotes the cone slant height. This quantity serves as a sensitive indicator of the breathing motion and collapse of the separated flow region. A similar numerical diagnostic approach was adopted in previous studies to assess the impact of free-stream enthalpy on shock-boundary layer interactions in double-wedge configurations~\citep{Reinert2017}.

\begin{figure}
    \centering
    \includegraphics[width=0.8\linewidth]{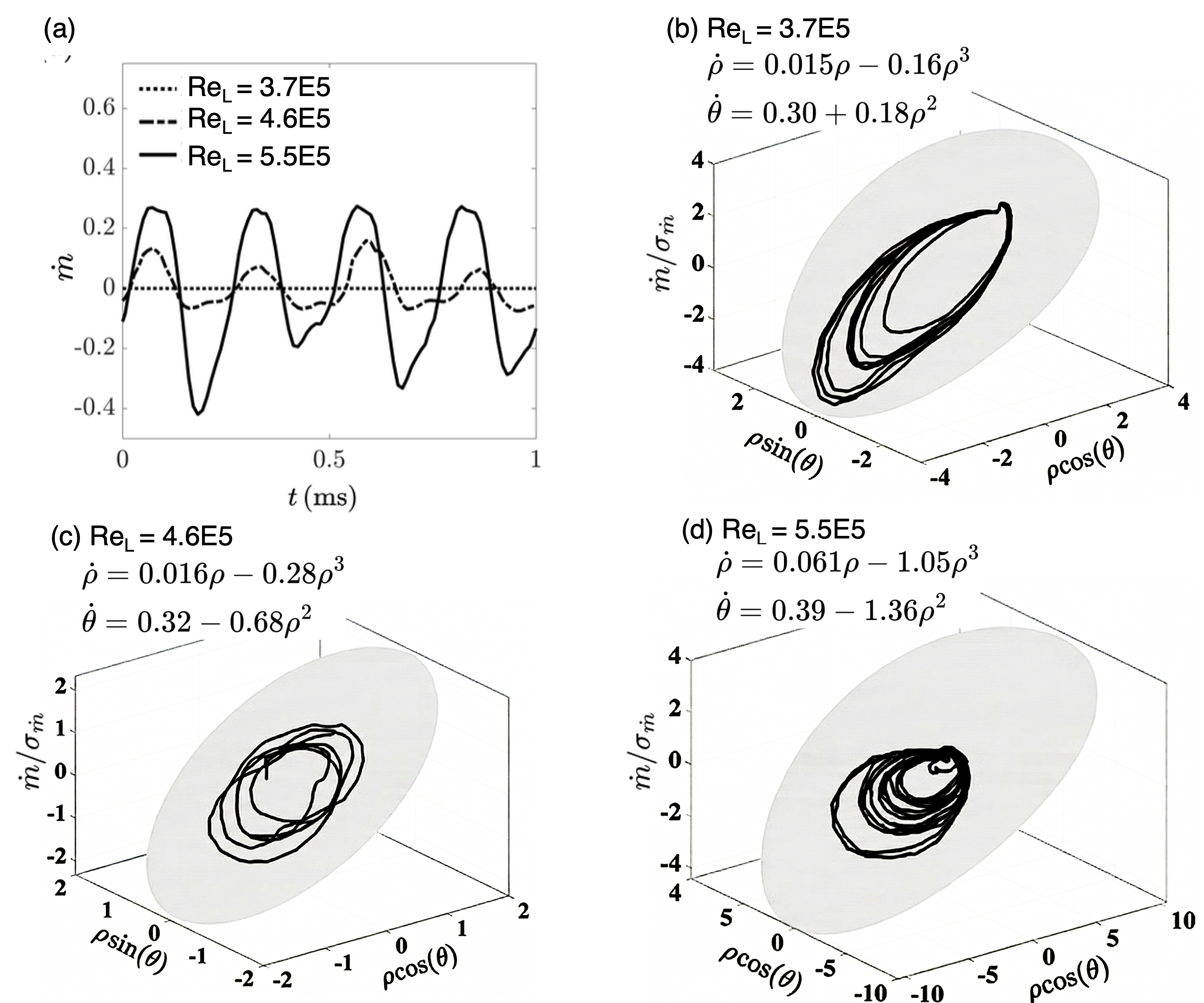}
    \caption{(a) Vertically-averaged mass flux  $  \dot{m}$ fluctuations at $L/2$ along the interaction region; (b)-(d) Spectral sub manifolds (SSM) and the reduced order nonlinear dynamics associated with the normalized mass flux as function of the freestream Reynolds number \textcolor{blue}{Re$_L$.}}
    \label{fig:phase_portrait}
\end{figure}

Representative time-series of the vertical plane mass flux at different \textcolor{blue}{Re$_L$} are shown in Figure~\ref{fig:phase_portrait} (a). In the following we utilize the spectral submanifolds (SSMs) for nonlinear model reduction using $\dot{m}(t)$. We apply delay embedding to construct a state vector $ y(t) \in \mathbb{R}^p $ of the form
\begin{align}
y(t) = [\dot{m}(t), \dot{m}(t+\tau), \dot{m}(t+2\tau), \dots, \dot{m}(t+(p-1)\tau)]^\mathrm{T},
\end{align}
where $ \tau $ is the time delay and $p \geq 2d + 1$  for a $ d$-dimensional SSM. We approximate the geometry of the attracting SSM by a third-order polynomial parameterization, i.e. $d=3$ and 
\begin{align}
\label{eq:SSMmod}
    y \approx v(\xi) = V\xi + M_2 \xi^{(2)} + M_3 \xi^{(3)},
\end{align}
where $\xi \in \mathbb{R}^2$ are the reduced coordinates,  $V \in \mathbb{R}^{p \times 2}$  spans the tangent space of the SSM, and  $M_2, M_3$ encode the curvature of the manifold. The reduced dynamics on the SSM is obtained as a third-order polynomial vector field~\citep{Szalai2017}
\begin{align}
\label{eq:dynmodSSM}
    \dot{\xi} = R\xi + R_2 \xi^{(2)} + R_3 \xi^{(3)},
\end{align}
where $R$  captures the linearized dynamics on the SSM, whereas the $R_2, R_3$ correspond to higher order terms of the dynamics on the third order SSM. Representation of the SSM parameterized by $(V,M_1, M_3)$ and the corresponding dynamics on it parameterized by $(R,R_1, R_3)$ are obtained by solving a constrained least squares problem using SSMLearn~\citep{Cenedese2022}.

{To capture nonlinear effects such as amplitude-dependent damping and frequency shifts, we transform the reduced dynamics into third-order complex normal form coordinates $z \in \mathbb{C}$. This involves projecting the real coordinates $\xi \in \mathbb{R}^2$ onto the eigenbasis of the linear part $R$ via a near-identity transformation of the form $\xi = W z + \mathcal{O}(|z|^2)$, where $R = W \Lambda W^{-1}$, $W$ contains eigenvectors, and $\Lambda$ is diagonal. This aligns the system with rotational eigendirections, enabling a polar representation $z = \rho e^{i\theta}$. For technical details, see~\citet{Szalai2017}.
%To capture the nonlinear behavior such as amplitude-dependent damping and frequency shifts associated with the identified dynamics, we transform the reduced model into its third-order complex normal form coordinates $z \in \mathbb{C}$. In polar coordinates $z = \rho e^{i\theta}$, 
The normal form of the dynamics then becomes}
\begin{align}
\label{eq:normal_form}
\dot{\rho} = \gamma(\rho)\rho = (\alpha_1 + \alpha_2 \rho^2)\rho, \quad
\dot{\theta} = \omega(\rho) = \beta_1 + \beta_2 \rho^2.
\end{align}
The third-order normal form is useful for identifying a supercritical Hopf bifurcation~\citep{Guckenheimer2013}, characterized by a change in sign of the linear damping coefficient $\alpha_1$. When $\alpha_1$ crosses zero from negative to positive, a stable limit cycle emerges if $\alpha_2 < 0$, indicating a supercritical Hopf bifurcation. This makes the third-order SSM a minimal yet sufficient model to capture nonlinear oscillatory behavior and detect the onset of such periodic motion in dissipative systems.

Figure~\ref{fig:phase_portrait} (b-d) shows the phase portrait of the non-dimensionalized mass flux normalized by its variance $\dot{m}/\sigma_{\dot{m}}$. Non-dimensionalization is performed by its mean value, and the trajectories are overlaid with the third-order SSM surface for increasing \textcolor{blue}{ Re$_L$}. The emergence of closed-loop trajectories indicates the presence of a stable limit cycle. 
The associated normal form dynamics derived from the third-order SSMs that best fit the observed time series data are also shown in the figure. 
%Insets in Figure~\ref{fig:phase_portrait}(b)–(d) display the corresponding normal form dynamics derived from the third-order SSMs that best fit the observed time series data. 
Note that irrespective of the \textcolor{blue}{Re$_L$}, the nonlinear coefficient $\alpha_2$ in the obtained normal form of the mass flux dynamics remains negative while the linear coefficient is positive $\alpha_1>0$, consistent with a limit cycle observed after a supercritical Hopf bifurcation. 

% \begin{figure}
%     \centering
%     \includegraphics[width=1.0\linewidth]{phasep_oscl_msflx.png}
%         \caption{Spectral sub manifolds (SSM) and the reduced order nonlinear dynamics associated with the normalized mass flux as function of the freestream Reynolds number \textcolor{blue}{Re$_L$.}}
%                 \label{fig:phase_portrait}
% \end{figure}
In the oscillatory regime, we observe that the linear coefficients $(\alpha_1, \beta_1)$ associated with amplitude and frequency dynamics remain nearly constant for \textcolor{blue}{Re$_L$ 3.7$\times10^5$ and 4.6$\times10^5$ cases.} This suggests the presence of a linear mechanism near the onset of bifurcation. With increasing Reynolds number, the nonlinear coefficients $(\alpha_2, \beta_2)$ increase in magnitude, indicating growing nonlinear effects in the system dynamics.

% At the highest Reynolds number ($\mathrm{Re} = 15.0\text{M}$), the nonlinear terms become comparable to or larger than the linear terms. This indicates that the truncated polynomial form of the spectral submanifold (SSM) model, using only cubic and quadratic terms, is insufficient to fully capture the flow dynamics at this Reynolds number. The stronger nonlinearity is evident in both the coefficient growth, the geometric deformation of the limit cycle in phase space (as seen in the SSM projections), as well as the time series of $\dot{m}(t)$ in Figure~\ref{fig:mass_flux_timeseries}. 

More specifically, as \textcolor{blue}{Re$_L$ increases from 3.7$\times10^5$ and 4.6$\times10^5$}, the growth in the linear coefficient $\alpha_1$ leads to a larger oscillation amplitude, while the growth in the nonlinear term $\alpha_2$ increases strength of the attractor leading to faster lock-in to the nonlinear limit cycle for the higher \textcolor{blue}{Reynolds number} case.

In addition, we also note a change in sign of the phase nonlinearity coefficient $\beta_2$. It changes from positive at \textcolor{blue}{3.7$\times10^5$ to negative at 4.6$\times10^5$ and 5.5$\times10^5$.} This change reflects a qualitative shift in how amplitude affects frequency, consistent with the experimentally and numerically observed frequency decrease (see Figure~\ref{fig:spod_cfd_exp_comp}).
However, with increasing nonlinearity, extending the SSM model to include higher-order terms becomes necessary to model the strongly nonlinear regime.

\section{Conclusions}
%\adw{R2-11: make conclusions more broad for a wider class of large cone setups} 
We study the emergence and evolution of unsteadiness in hypersonic shock-boundary layer interactions over a cone-step geometry. Through a combination of quiet wind tunnel experiments and high-fidelity numerical simulations, the study demonstrates the role of \textcolor{blue}{freestream Reynolds number} as an additional control parameter alongside geometric features (cone-length to base diameter ratio and the cone half angle) that govern the flow stability and unsteadiness. %Despite the specific choice of the cone step geometry, \upd{the conclusions from our study apply to a broader class of large angle double cone configurations and axisymmetric inlet instabilities which report a similar unsteadiness signature}

The key findings of our study are as follows:
\begin{itemize}
    \item We identify distinct regimes of unsteadiness as the Reynolds number increases. The regimes range from (i) fluctuations due to shear-layer breakdown, (ii) small-amplitude oscillations in the separation shock system and (iii) large-amplitude oscillations in the separation shock system resulting in a pulsatory motion and transitory detachment of the bow shock from the cone tip.
    \item SPOD analysis shows that the dominant unsteady mode in the small-amplitude regime is associated with the motion of the separation and bow shock and has a Strouhal number of approximately St $\approx 0.17$. For the highest \textcolor{blue}{Re$_L$} case considered, the Strouhal number decreases to St = 0.13.
    \item The simulations help characterize the unsteady oscillations of the shock system in the quasi-linear small-amplitude and the pulsatory large-amplitude  regimes.   
    %\textcolor{red}{qualify this better; see sections in the paper which may help communicate this point better.}
    \item The role of the contact discontinuity on the separated shear layer instability is quantified via a novel one-dimensional stability analysis. The contact emanates from the nose tip and the separation shock interaction and impinges on the shear layer causing perturbation confinement and rapid breakdown.
    \item Nonlinear model reduction using spectral submanifolds (SSMs) suggests that the oscillations result from a supercritical Hopf bifurcation. Additionally, the azimuthal velocity dynamics in the recirculation resembles global linear modes analyzed previously in the literature. This suggests a hydrodynamic origin of the instability.
\end{itemize}

Future work will entail further investigation of the onset and origin of the unsteadiness across the geometric space. \upd{The conclusions from our study apply to a broader class of large angle double cone configurations and axisymmetric inlet instabilities which report a similar unsteadiness signature.} 

% Key findings include the identification of  The onset of unsteadiness correlates strongly with the interaction between the contact discontinuity and the separation shear layer, a process found to enhance instability via a confinement effect. SPOD analyses confirm that the dominant unsteady mode has a hydrodynamic origin, characterized by a consistent Strouhal number of approximately $St \approx 0.17$ across a range of conditions. 

% Furthermore, nonlinear model reduction using spectral submanifolds (SSMs) reveals that the transition to periodic unsteadiness follows a supercritical Hopf bifurcation. This insight quantitatively supports the observed progression of flow dynamics from linear to nonlinear regimes. The study's combined experimental and computational approach provides critical understanding and predictive capability for the design of hypersonic vehicles encountering SBLI phenomena. Future work could explore the effects of external disturbances, vibrational non-equilibrium, and more complex geometries to further generalize the insights obtained here.

\section*{Supplementary Material}
Supplementary movie associated with this manuscript is available presently in the JFM review system. 

\section*{Acknowledgments}
Author S.G.S thanks Bhavith Sai for help with data collation in Figure 3.
% \section*{Funding}
Authors C.J., E.L.C., and J.S.J. were supported by a gift from Northrop Grumman Corporation and Purdue University School of Aeronautics and Astronautics AAE520 laboratory class funds. Author C.J. was additionally supported by a National Science Foundation Graduate Research Fellowship. 

\section*{Declaration of Interests}
The authors report no conflict of interest.

% \section{Scratch}
% \begin{align}
%     \theta = 25^\circ \\
%     \theta = 30^\circ \\
%     \theta = 35^\circ \\
%     \text{Re}_x = 12.5\text{M} \\
%     \text{Re}_x = 15.0\text{M} \\
% \end{align}

% \section{Scratch}

\appendix
%\section{Comparison of present work in the unsteadiness parameter space     }
{\color{blue}
\section{Influence of contact discontinuity on shear layer stability}  
\label{sec:lst_contact}

In flow configurations considered in this study, early separation leads to the cone-tip and separation shock interaction near the cone surface. This interaction generates a contact discontinuity that impinges on the separated shear layer downstream.
The contact-shear layer interaction only occurs for certain geometries. For example, in configurations with lower $\theta_1$, the interaction may occur far downstream or may not occur at all. However, when the interaction occurs close to the separation point, rapid shear layer destabilization is observed, characterized by strong unsteadiness in both simulations and experiments.
For the present cases, this interaction is found to correlate with the spatial onset of shear layer unsteadiness. We hypothesize that the proximity of the contact discontinuity to the shear layer plays an important role in triggering this destabilization. To examine this hypothesis, a one-dimensional (1D) compressible linear stability analysis (LSA) is carried out.

For asymptotic stability analysis, the temporal form of the perturbations is assumed to be exponential. For the linear stability analysis, we assume that perturbations are two-dimensional and that they are periodic in the streamwise direction $x_1$ and can be expressed as,
\begin{align}
\mathbf{U}^{\prime}\left(x_j, t\right)\;=\;\mathbf{U}^{\prime}\left(x_2\right) \, \exp \left(-i\left(\omega_r+i \omega_i\right) t+i k_x x_1\right).
\end{align}
The  governing equations with the perturbations lead to an eigenvalue problem. The temporal growth rate of the perturbations is given by $\omega_i$ and linear instability exists when $\omega_i>0$.

% \subsubsection{\bf 1D LSA of Shear Layer in the vicinity of a Contact Discontinuity}

\begin{figure}
    \centering
    \includegraphics[width=1\linewidth]{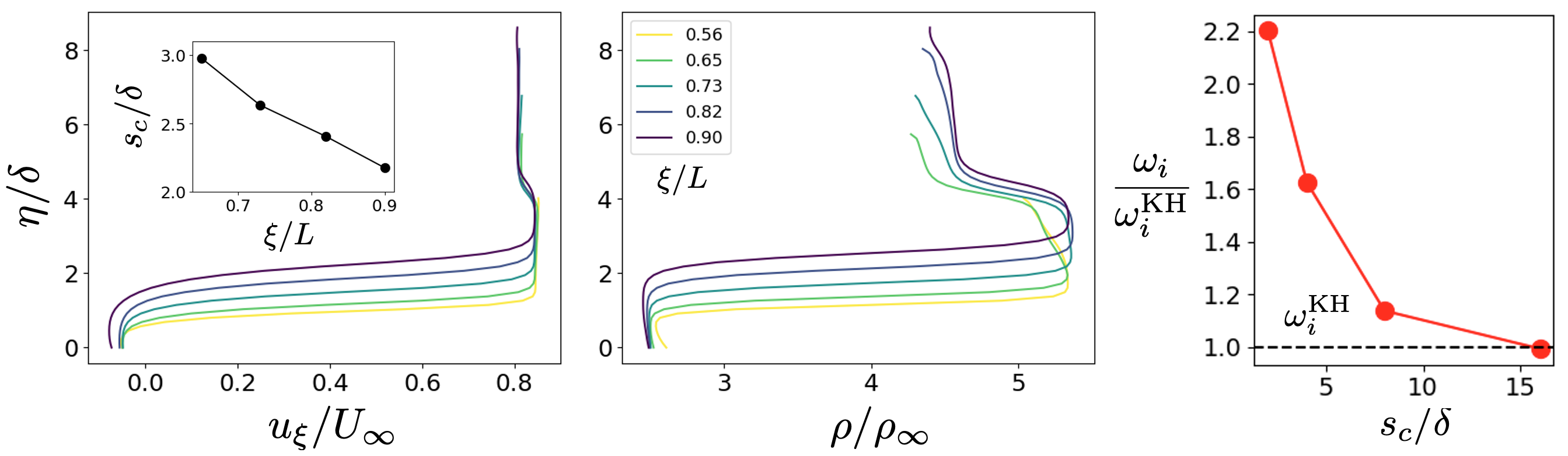}
    \caption{Base flow profiles from the 2D axisymmetric flow (case $\theta_1=25^\circ$, \textcolor{blue}{Re$_L$}=7.5$\times10^5$) for (a) streamwise velocity, (b) density, and (c) the ratio of the growth rate from the stability analysis with and without the contact discontinuity.}
    \label{fig:base_prof1dlst}
\end{figure}

\begin{figure}
    \centering
    \includegraphics[width=0.9\linewidth]{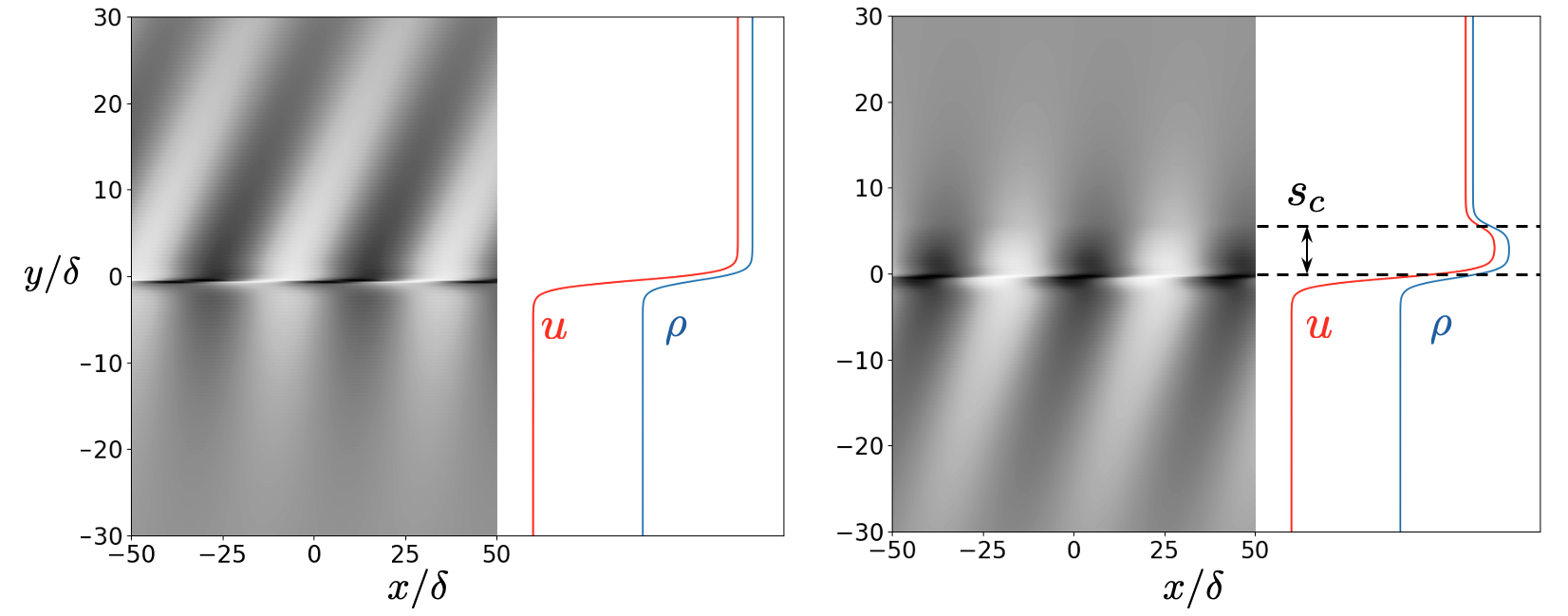}
    \caption{ Density perturbations associated with the most unstable mode with $\alpha$ = 0.4, corresponding to base flow profiles (a)  without contact discontinuity (SL), and (b) with contact discontinuity (SLC).}
    \label{fig:mode_shapeLST}
\end{figure}
The stability analysis of the 2D axisymmetric flow profiles extracted perpendicular to the shear layer from our simulations is shown in Figure~\ref{fig:base_prof1dlst}. Two primary base flow configurations are analyzed:
\begin{enumerate}
    \item A baseline profile with only the shear layer (SL) of thickness $\delta$.
    \item A composite profile with a contact discontinuity of the same thickness immediately adjacent to the shear layer, as is the case in the cone-step flow. 
\end{enumerate}

The distance between the shear layer and the contact discontinuity $s_c$ is varied from $2 \delta$ to $16 \delta$.
% The stability analysis involves solving the linearized governing equations (e.g., Euler or Navier-Stokes) for small-amplitude perturbations superimposed onto the chosen base flow profile. This allows for the determination of the amplification rates of different disturbance frequencies and wavenumbers. 
The results from the LSA are presented in terms of the maximum spatial growth rate ($\omega_i$) of the most unstable disturbance mode $k_x \delta =0.4$ as a function of distance $s_c/\delta$ between the shear layer and the contact discontinuity. This is shown in Figure~\ref{fig:base_prof1dlst} (c) in the SLC profile. Note that as $s_c/\delta  \gg 1$, we recover the results corresponding to the SL profile. 

The analysis indicates that the growth rate of the instability associated with the SLC profile is approximately twice that of the baseline SL profile and will grow quadratically as fast as the instabilities in an isolated SL profile. The modes from the stability analysis are shown in Figure~\ref{fig:mode_shapeLST} and shed light on the physical mechanism associated with the additional growth. The instability in the baseline SL profile corresponds to the classical Kelvin-Helmholtz mechanism. The enhancement of growth rate for the SLC case occurs due to the confinement effect provided by the contact on the faster moving fluid. This confirms that the interaction between the contact discontinuity and the shear layer acts as a powerful destabilizing mechanism, amplifying disturbances much more effectively than the shear layer alone, and therefore, it likely plays a critical role in accelerating the onset of shear layer breakdown and the consequent appearance of small-scale fluctuations.
}

\section{Predictions from an acoustic resonance model for the oscillation frequency}
% \textcolor{red}{change the numbers based on latest calculations}
In this section, we consider an empirical 
 acoustic resonance model, recently proposed in \cite{Kumar_Sasidharan_Kumara_Duvvuri_2024}, which uses a Rossiter-type {\it cavity} resonance expression. The model and its applicability are tested on the experimental data acquired in this work. We briefly discuss the model assumptions, its expression for the oscillation frequency, and compare it against our experimental data.

The model hypothesizes an acoustic feedback loop. This hypothesis stems from the similarities between the cone step flow and compressible open-cavity flows. The latter, as discussed in the literature review, are known to exhibit oscillations due to an acoustic feedback mechanism. Both flows feature a separation bubble and a compressible shear layer that impinges downstream. Based on this similarity, an acoustic model was proposed by the authors to predict the oscillation frequency.

The model proposes the following feedback loop: (i) downstream propagation of shear layer disturbances at convective speed $u_c$ and (ii) upstream propagation of acoustic waves generated by the impinging shear layer, through the subsonic separated region.

It should be noted that the cone step flow considered in this work does not satisfy the Rossiter's geometric criterion for existence of acoustic resonance via standing waves. The equivalent cavity length-to-depth ratio is greater than 4.  Nevertheless, we apply the model to the oscillatory case presented in this paper and demonstrate that it fails to predict the Strouhal number. This discrepancy can be attributed to several inconsistent physical assumptions within the model when applied to oscillatory shock-boundary layer interactions.

\par The acoustic resonance model predicts the m$^\text{th}$ mode Strouhal number as
\begin{equation}
St = \frac{fL}{U_\infty} = \left(\frac{1}{M_\infty}\right) \left(\frac{a_o}{a_\infty}\right) \left(\frac{l_1}{L_s}\right) \left(\frac{m-\kappa}{\frac{1}{KM_o} + \frac{a_o}{a_i}}\right)
\end{equation}

Here $i$ and $o$, as in the original model, refer to the low and the high-speed sides of the shear layer and $L_s$ is the distance from the separation point to the reattachment point of the shear layer. The ratio $r=a_o/a_i$ can be calculated using the assumptions about the propagation speed of the acoustic wave. 
%The equation below was used to calculate $\frac{a_o}{a_i}$: \[r=\frac{1}{\left[1+\sqrt{Pr}\left(\frac{\gamma-1}{2}\right)M_o^2\right]^{0.5}}\] 
$K$ is defined as the ratio $u_c/u_o$, which depends on the
acoustic wave of interest. In this model, acoustic waves generated inside the `cavity' are relevant. This corresponds to the Kelvin-Helmholtz waves in \cite{OertelSen2016} model. 
%type of `choice' of disturbance in the shear layer and is determined empirically.
% The three equations below were used to calculate $K$ for the three different disturbance modes:

% \[K= \begin{cases} 
%       \frac{1+2r}{(1+r)^2} & K-H \hspace{1cm} mode \\
%       \frac{1}{1+r} & Supersonic \hfill mode \\
%       \frac{1}{(1+r)^2} & Subsonic \hfill mode 
%    \end{cases}
%\]
%The ratio $\frac{a_o}{a_\infty}$ was calculated using the equation below:
%\[\frac{a_o}{a_\infty} = \left[\frac{1+(\frac{\gamma-1}{2})M_\infty^2}{1+(\frac{\gamma-1}{2})M_o^2}\right]^{0.5}\]
 $\kappa$ is the phase difference between the downstream propagating disturbances in the shear layer and $m$ is the mode of oscillation. $m$ is set to 1 for the first mode of oscillation. The value of $\kappa$ is chosen to be 0.25 as used in \cite{rossiter1964report}.
 For the case Re 10M and $\theta_1=35^\circ$, the model predicts a subsonic acoustic disturbance. Using $K= (1+2r)/(1+r)^2$ from \citet{OertelSen2016}, we obtain a predicted frequency of $6156$ Hz as shown in table~\ref{tab:freq_comparison}.

\begin{table}
\centering
\begin{tabular}{c c}
\hline
 \textbf{Predicted (Hz)} & \textbf{Measured (Hz)} \\
\hline
6156 & 3900 \\
\hline
\end{tabular}
\caption{\cite{Kumar_Sasidharan_Kumara_Duvvuri_2024} acoustic resonance model prediction for $\theta_1=35^\circ$, Re=10M oscillatory case. }
\label{tab:freq_comparison}
\end{table}

\par As is seen, the model prediction error is approximately 58\%. This is because of the inapplicability of the acoustic resonance model in the cone-step flow. Our  results suggest a hydrodynamic phenomenon and future research is warranted. 
%The assumption that the flow velocity is zero beneath the shear layer is only true at the center of the separation bubble, and it can't be assumed that the flow beneath the shear layer is stagnant. 
Additionally, the model also fails to take into account the contact above the shear layer, which emanates from the intersection of two shocks impinges on the shear layer.
%dividing the region above the shear layer into two regions and making the high-speed side properties ambiguous. 

%WE NEED TO PUT THE MEASURED VS PREDICTED FREQUENCY (OR STROUHAL NUMBER)

%\section{Effect of Re and $\theta_1$ on the contact-shear layer interaction}

{\color{blue}

\section{Grid and azimuthal sector convergence}

We demonstrate that the results are insensitive to both grid resolution and domain extent. In this context, domain convergence refers to the azimuthal sector size employed in the simulations. Two sector extents, \(36^\circ\) and \(72^\circ\), were examined to assess whether the azimuthal domain influences the resolved flow features. The comparison shows that the \(72^\circ\) sector adequately captures all dominant azimuthal wavenumbers associated with the unsteady flow structures in the separation and reattachment regions. Hence, this extent is deemed sufficient for converged spectral and statistical results.

Details of the grid resolution are summarized as follows. The fine-sector grid is designed to ensure adequate wall-normal resolution for both the boundary layer and the separated shear layer. On the cone surface, the grid employs a first-cell height of \(\Delta y =\) 0.001 mm corresponding to \(y^+ < 1\) everywhere on the cone surface. In the shear layer, the normal resolution is approximately \(\Delta y =\) 0.02 mm. The streamwise grid spacing near the separating boundary layer is approximately \(\Delta x = 0.07\) mm, and the azimuthal resolution in the entire domain is given by \(\Delta \theta = 0.56^\circ\). Three progressively refined grids: coarse, medium, and fine are used for the $72^\circ$ sector. The coarse grid for contains approximately 10 million grid points, the medium grid roughly 16 million, and the fine grid around 25 million points.

As shown in Fig \ref{fig:grid_conv} (a), the pressure coefficient (\(C_p\)) distributions exhibit convergence with increasing grid resolution, confirming adequate spatial accuracy in the presence of strong and multiscale pulsating unsteadiness. The mean \(C_p\) field serves as a stringent measure of solution convergence in presence of large pressure variations that exceed the mean value.

Furthermore, Fig. \ref{fig:grid_conv} (b) presents the convergence of the SPOD leading-mode energy spectra for the different grid resolutions. The results demonstrate a consistent spectral peak, with the dominant frequency near \(f \approx 0.14\) remaining unaffected by further refinement, reinforcing the robustness of the chosen resolution and domain size.

\begin{figure}
    \centering
    \includegraphics[width=1\linewidth]{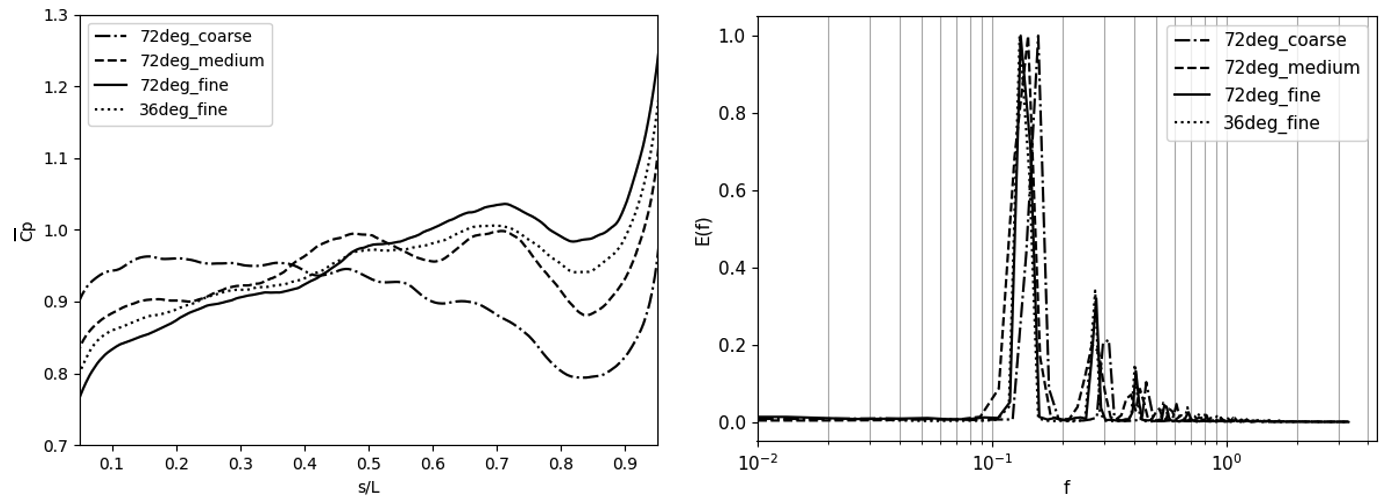}
    \caption{Grid and azimuthal domain extent effect on (a) mean pressure on the cone and (b) the leading mode SPOD energy spectrum for the largest Reynolds number case Re$_L=5.5\times 10^5$.}
    \label{fig:grid_conv}
\end{figure}

\section{SPOD analysis details}
\label{spod_details}
\noindent
{\bf Experimental schlieren data}

The SPOD analysis is performed directly on the intensity fluctuation field obtained from schlieren images. 
%Since only a single observable is available, a unit weighting matrix (identity) is employed, and no inverse reconstruction of the density field is carried out.
The sampling frequency is equal to the camera frame rate, \( f_s = 78~\text{kHz} \), with a total of 1000 frames used for analysis. This corresponds to a time window of \( 12.8~\text{ms} \) (dimensional) and approximately 295 non-dimensional time units. The resolvable frequency range extends from \( 78~\text{Hz} \) to \( 39~\text{kHz} \), corresponding to non-dimensional frequencies of \( 3.4\times10^{-4} \) to \( 1.695 \). 
%Convergence of the leading-mode energy with respect to the sampling rate is verified, as shown in Fig.~\ref{fig:spod_conv_sampling}.
Hamming windows are applied in the spectral estimation, with variations in block length and overlap examined to assess convergence and spatial coherence of the dominant modes. 
%Representative results for the \(\text{Re}/m = 15M\) case are shown in Fig.~\ref{fig:exp_blk_ovlp}, illustrating stable convergence behavior for multiple window configurations.
% \begin{figure}[h!]
%     \centering
%     \includegraphics[width=0.5\linewidth]{revised_figs/exp_blk_ovlp.png}
%     \caption{Convergence of the SPOD modal energy for various block window lengths and overlaps for  Re/m = 15M case. }
%     \label{fig:exp_blk_ovlp}
% \end{figure}

Analysis of suboptimal modes indicates that the leading SPOD mode consistently contains fluctuation energy several orders of magnitude higher than the subsequent modes. Hence, the analysis primarily focuses on the dominant frequency associated with this leading mode. \\

\noindent
{\bf Simulation sector plane data}

The SPOD of the planar simulation data is performed using a sampling frequency of \( f_s = 166.67~\text{kHz} \). The dataset includes the normalized flow variables \( \rho, u_i, \text{ and } T \). An analysis using only the density field \( (\rho) \) produces nearly identical dominant frequency peaks.
A total of 1000 frames are used, corresponding to a total dimensional time of \( 6.6~\text{ms} \) and about 140 non-dimensional time units. The minimum and maximum resolvable frequencies are \( 0.167~\text{kHz} \) and \( 83.3~\text{kHz} \), respectively, yielding a non-dimensional frequency range of \( 7.2\times10^{-3} \) to \( 3.621 \). 
%Convergence of the leading mode energy with respect to sampling rate is verified, as illustrated in Fig.~\ref{fig:placeholder}.

}

\bibliographystyle{jfm}
% % Note the spaces between the initials
\bibliography{jfm-instructions}

\end{document}